\documentclass[
    11pt,
    letterpaper,
    reprint,
    notitlepage,
    superscriptaddress,
    aps,
    prl
]{revtex4-2}

\usepackage[dvipsnames, usenames]{xcolor}
\usepackage{graphicx}
\usepackage{bm}
\usepackage{physics}
\usepackage{mathtools}
\usepackage{soul}
\usepackage[normalem]{ulem}

\usepackage{amsmath,amssymb,amsfonts} 
\usepackage[hypertexnames=false]{hyperref}
\usepackage[capitalize]{cleveref}
\usepackage{tikz}
\usetikzlibrary{quantikz}
\hypersetup{colorlinks=true,citecolor=blue,linkcolor=blue,urlcolor=blue}

\begin{document}

\title{Extending the dynamic range in quantum frequency estimation with sequential weak measurements}

\author{Su Direkci}
\affiliation{California Institute of Technology, Pasadena, CA 91125, USA}
\author{Manuel Endres}
\affiliation{California Institute of Technology, Pasadena, CA 91125, USA}
\author{Tuvia Gefen}
\affiliation{Racah Institute of Physics, The Hebrew University of Jerusalem, Jerusalem 91904, Givat Ram, Israel}

\date{\today}

\begin{abstract}
    Quantum metrology explores optimal quantum protocols for parameter estimation. In the context of optical atomic clocks, conventional protocols focus on  optimal input states and measurements to achieve enhanced sensitivities. However, such protocols are typically limited by phase slip errors inflicted due to the decoherence of the local oscillator. 
    Here, we study schemes to 
    extend the dynamic range and overcome phase slip noise through weak measurements with ancilla qubits. Using coherent spin states, we
    find optimal weak measurements protocols: we identify optimal measurement strength for any given interrogation time and number of atoms.
    Then, we combine weak and projective measurements to construct a protocol that asymptotically saturates the noiseless precision limits, and outperforms
    previously proposed methods for phase-slip noise suppression.
\end{abstract}

\maketitle

In quantum frequency estimation, we seek to estimate an unknown frequency $\omega$, given a dynamic range of $[\omega_0, \omega_0+\delta \omega],$ with an optimal
precision.
Maintaining a high precision given an arbitrarily large frequency bandwidth, $\delta \omega,$ is one of the fundamental challenges 
in quantum metrology, and it is typically referred to as the precision-bandwidth tradeoff \cite{rosenband2013exponential,kaubruegger_quantum_2021, marciniak2022optimal,direkci2024heisenberg,liu2025enhancingdynamicrangesubquantumlimit}.
The standard estimation schemes rely on a Ramsey-type experiment: $N$ qubits are initialized to an input state, rotate at the frequency $\omega$ for time $T$, and are subsequently measured.
For an infinitesimally small bandwidth, longer evolution times enhance the precision, leading to estimation uncertainty that scales as $1/T$, known as Heisenberg scaling (HS) with respect to time \cite{degen_review, higgins_entanglement-free_2007,
de2005quantum, adaptive_bayesian_clock,huang2023learning}. However, a finite bandwidth restricts this improvement: an evolution time that is longer than $\pi/\delta \omega$ 
leads to phase slip errors that considerably degrade the precision. Consequently, the evolution time in systems such as optical atomic clocks \cite{bloom2014optical, Ludlow_opt_atomic_clocks, oelker2019demonstration, Norcia2019,Majdarov2019,robinson2024direct} is typically constrained by the inverse of the frequency bandwidth \cite{Leroux_2017}.
This phenomenon is prominent for the Greenberger–Horne–Zeilinger (GHZ) state, as it provides optimal sensitivity but suffers from a poor dynamic range, limiting its usefulness when $\delta\omega T$ is large \cite{berry_how_2009,macieszczak_bayesian_2014, jarzyna_true_2015, kaubruegger_quantum_2021}. 
It is therefore highly desirable to develop schemes that suppress phase slip errors and achieve nearly optimal precision for a large bandwidth.

\begin{figure}[h]
    \centering   
\includegraphics[width=0.9\linewidth]{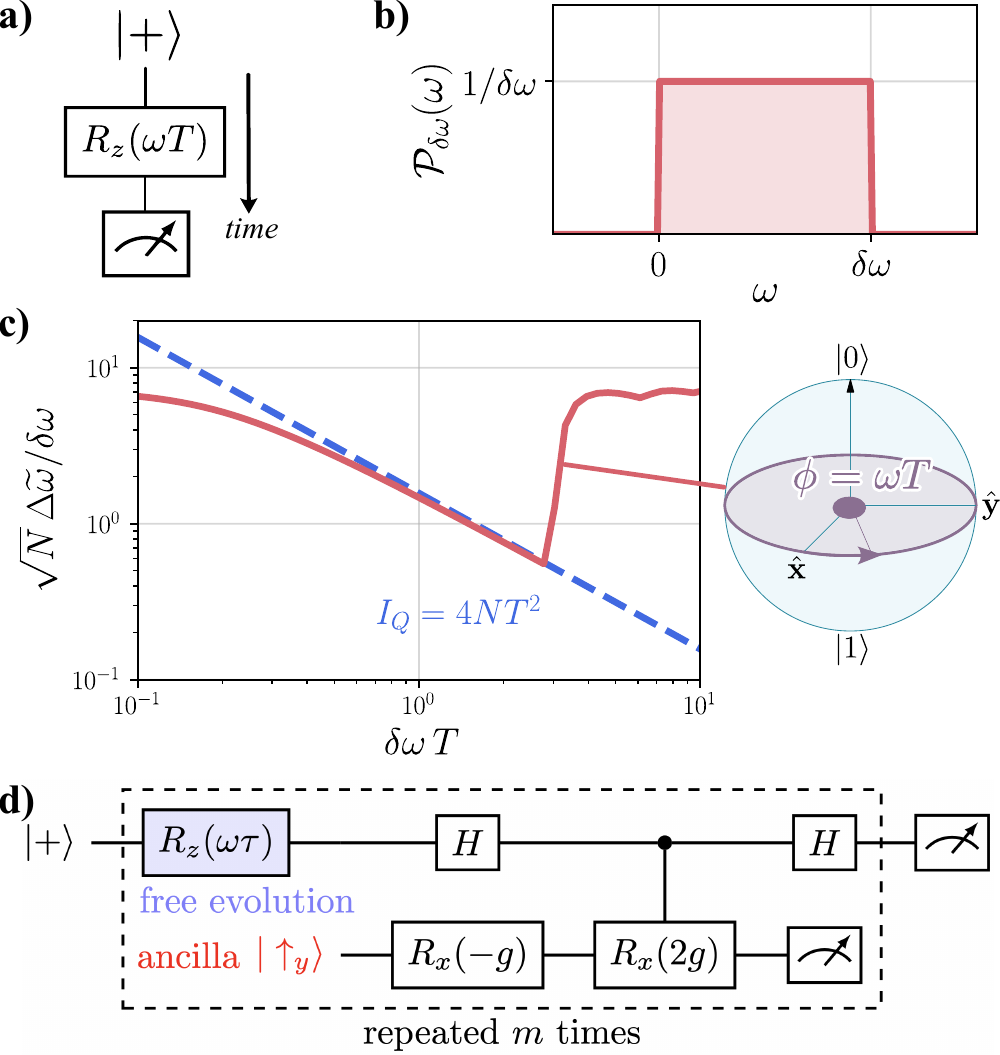}
\caption{Dynamic range problem in frequency estimation: a) Circuit representation of the standard Ramsey experiment, where a strong measurement is applied after a phase accumulation time $T.$
b) We consider a uniform prior distribution of the unknown frequency $\mathcal{P}_{\delta\omega}(\omega)$ in the interval $[0, \delta\omega]$. c) The square-root of the Bayesian mean squared error (BMSE) $\Delta\widetilde{\omega}$, normalized by the prior width $\delta\omega$ and the number of qubits $N=64$. The solid, red line corresponds to the measurement-optimized BMSE, which suffers from phase slips beyond $\delta\omega\,T > \pi$, as shown in the Bloch sphere picture. The dashed-blue line corresponds to the quantum Fisher information bound of $(\Delta\omega)^{-2}=I_{Q}=4NT^2.$ 
d) Circuit representation of the approach studied here: Ramsey experiment with sequential weak measurements. }
\label{fig:summary_fig}
\end{figure}

Several schemes successfully combine entanglement enhanced sensing with a relatively large dynamic range, i.e. $\delta \omega \approx \pi/ T$ \cite{berry_how_2009,macieszczak_bayesian_2014, kaubruegger_quantum_2021,direkci2024heisenberg,kessler_heisenberg-limited_2014, pezze_smerzi_HybridCoherentSqueezed}.
However, in most practical applications it is important to extend the dynamic range even further: for example, in atomic clocks, 
larger dynamic range
would improve the tolerance to laser noise and the clock stability. 
The fundamental question is, therefore, how to extend the dynamic range beyond this limit. 
Several methods have been proposed for this purpose, including cascaded protocols \cite{rosenband2013exponential, kessler_heisenberg-limited_2014, direkci2024heisenberg}, quantum deamplification \cite{liu2025enhancingdynamicrangesubquantumlimit},
and sequential weak measurements \cite{shiga_locking_2012, borregaard_near-heisenberg-limited_2013}.
However, it is unknown whether these schemes saturate the relevant precision bounds, and it is highly desirable to find schemes that are both optimal and easy to implement with state-of-the-art platforms. 

We focus in this Letter on the weak measurement approach. While sequential weak measurements were proposed in the context of cavity-based atomic clocks \cite{shiga_locking_2012, borregaard_near-heisenberg-limited_2013},
there has been no systematic study of their precision limits in the context of Bayesian frequency estimation. In particular, it is unknown whether such schemes achieve HS or saturate the precision bounds for a large bandwidth, especially for a limited number of atoms (hereafter,
HS refers only to the scaling with time).
Lastly, prior works assumed a large ensemble of atoms measured collectively \cite{shiga_locking_2012, borregaard_near-heisenberg-limited_2013}; devising an optimal weak measurement protocol for circuit-based sensors \cite{shaw_multi-ensemble_2023, finkelstein2024universal, cao2024multiqubit} and for a small number of atoms \cite{sensing_with_single_qubit} remains an open challenge. 
We address these questions by designing a tunable sequential weak measurement protocol and analyze its performance for different configurations.
We focus on the simplest case where the input state is a coherent spin state (CSS), and propose a protocol that asymptotically saturates the ultimate precision bounds for CSS for an arbitrary large bandwidth.  This protocol also shows HS and outperforms conventional protocols in the finite $N$ regime.


{\textit{Formulation and Cram\'er-Rao bounds---}}We consider a Ramsey type experiment: $N$ qubits evolve according to the Hamiltonian $\mathcal{H}=\omega{\sum_{j=1}^N}\sigma_z^{j},$
where $\sigma_z^{j}$ is the Pauli $z$ operator acting on the $j^\text{th}$ qubit, and $\omega$ is the unknown frequency to be estimated.  
We consider the Bayesian setting where $\omega$ has a uniform prior distribution $\mathcal{P}_{\delta\omega}\left(\omega\right)=\frac{1}{\delta\omega}\chi_{\left[0,\delta\omega\right]}\left(\omega\right)$ in 
$[0,\delta\omega]$ ($\chi$ is the indicator function) \footnote{Extension to other forms of prior distributions is discussed in \cite{supp}.}. 
Note that any uniform prior distribution in the form of $\left[\omega_{0},\omega_{0}+\delta\omega\right]$ is equivalent to this distribution: we can retrieve the original prior by applying a unitary of $e^{i\omega_{0}t\sum_{j}\sigma_z^{j}}$ before the measurement. The error in the estimation is quantified by the Bayesian mean squared error (BMSE), which is defined as the mean squared error (MSE) weighted by the prior distribution: 
\begin{align}
\label{eqn:bmse_mse}
(\Delta \widetilde{\omega})^2 &= \int (\Delta \omega)^2 \mathcal{P}_{\delta\omega}(\omega) d\omega \,
\end{align}
where $(\Delta \omega)^2 := \sum_{x} p(x|\omega)(\omega-\omega_{\text{est}}(x))^2$ is the MSE,
$(\Delta \widetilde{\omega})^2$ 
is the BMSE, $\{p(x|\omega) \}_x$ are the probabilities of the measurement results, and  $\{\omega_{\text{est}}(x)\}$ is the estimator function. 

The standard Ramsey experiment is illustrated in Fig. \ref{fig:summary_fig}a. The probes are initialized to a CSS, $|+\rangle^{\otimes N}=(\ket{0} + \ket{1})^{\otimes N}/2^{N/2}$,
where $|+\rangle$ is an eigenstate of the Pauli $x$ operator. They are subjected to a free evolution for a duration of $T$, and are finally measured. The BMSE of this protocol, when optimized over all possible measurements, is referred to as the optimal classical interferometer (OCI) \cite{macieszczak_bayesian_2014,supp} 
 and is plotted in Fig. \ref{fig:summary_fig}c.
We observe that for interrogation times where $\delta\omega \, T > \pi$, the OCI BMSE increases rapidly, and converges to the prior width, $\Delta\widetilde{\omega} \approx \delta\omega$, signifying that no information is gained from the estimation. The behavior is caused by phase slips: the unitary transformation describing the free evolution is periodic in $\omega \,T$. Therefore,
the maximum interrogation time using this protocol is limited to $\pi/\delta \omega$, since longer times will give rise to phase slip errors.

To resolve this issue, we propose to perform sequential weak measurements during the free evolution.
Weak measurements provide frequent monitoring of the state and allow us to track the oscillations, preventing phase slip errors \cite{gefen_quantum_2018, cujia_tracking_2019, pfender_high-resolution_2019,cohen2020achieving, tratzmiller2020limited}. They therefore can potentially extend the dynamic range beyond $\pi/T.$  

Our scheme consists of applying a sequence of weak $\sigma_{x}$ measurements in time intervals of $\tau$.
They are described by the Kraus operators $K_{\pm} = \left( \cos{(g)} I \pm \sin{(g)} \sigma_{x} \right)/\sqrt{2},$ where $g$ is a tunable parameter that represents the measurement strength ($g=\pi/4$ corresponds to a projective measurement). They can be realized by weakly entangling each probe to an ancillary qubit and measuring the ancilla \cite{supp}.
Each step of the protocol involves the rotation $V_{\omega}=\cos\left(\omega\tau\right)I-i\sin\left(\omega\tau\right)\sigma_{z},$ followed by a weak measurement that transforms the state of the sensing qubit $\rho$ as $\rho \rightarrow K_{\pm} \, \rho \, K_{\pm}^\dagger/\text{Tr}[K_{\pm} \, \rho \, K_{\pm}^\dagger]$, depending on the measurement outcome. Therefore, the sensing qubits evolve in stochastic quantum trajectories \cite{ilias_biasing_2025}.
Given a total interrogation time $T$ and a measurement period $\tau$, this stochastic step is repeated for $m=\lfloor T/\tau \rfloor$ times \footnote{Note that if we set $\tau =T$, $g = \pi/4$, this protocol reduces to the standard Ramsey experiment.}. Note that, to prevent phase slips, the measurement period must satisfy $\tau <\frac{\pi}{2\delta\omega}$.
We study two different protocols, illustrated in Fig. \ref{fig:summary_fig}d: the \textit{weak-only} protocol, in which only sequential weak measurements are performed, and the \textit{weak-with-strong} protocol, in which the final weak measurement is replaced by a projective measurement on the sensing qubits.

Let us study the effects of weak measurements and the resulting stochastic dynamics.
Initializing the sensing qubit in the state $|+\rangle$,
it remains on the $\sigma_x$-$\sigma_y$ plane \footnote{We ignore decoherence effects, such as amplitude damping. See \cite{supp} for a derivation.}, 
and is thus fully characterized by the angle $\theta$ in this plane.
This angle is a random variable that depends on the previous measurement outcomes. It can be calculated iteratively using the following relation: $\theta_{k+1}=-2\omega\tau +\text{arg}\left(\cos\left(\theta_{k}\right)+(-1)^{x_{k}}\sin\left(2g\right)+i\cos\left(2g\right)\sin\left(\theta_{k}\right)\right)$,
where $\theta_k$ is the angle before the $k^\text{th}$ measurement, and $x_k=0,1,$ corresponds to the outcome of the $k^\text{th}$ measurement. Notice that
in the limit of $g\rightarrow 0,$ $\theta_{k+1} = \theta_k - 2\omega\tau$, which corresponds to the free evolution in a standard Ramsey experiment.
Each step thus consists of a rotation by an angle of $2 \omega \tau,$ followed by a small stochastic kick in the phase that  degrades the coherent phase accumulation. This effect can be seen more clearly from studying the average dynamics, corresponding to the sequential application of the channel $\Lambda\left(\rho\right)={\sum_{i\in\left\{ \pm\right\} }}K_{i}^{\left(\omega\right)}\rho K_{i}^{\left(\omega\right)\dagger}$, where $K_{i}^{\left(\omega\right)}=K_{i}V_{\omega}.$ 
The sensing qubit undergoes dephasing in this channel with a rate of $\gamma=-\frac{1}{2\tau}\log\left(\cos\left(2g\right)\right)\approx g^{2}/\tau$ in the limit of $g \ll 1$.
This dephasing stems from the phase random walk induced by the measurements, a phenomenon that is typically termed as \textit{measurement back action}.
Even though in our protocol we are not limited to the average dynamics but track individual trajectories,
we will show that this dephasing still affects the sensitivity.

We are now poised to study the precision limits of this scheme and compare them to the fundamental bounds. The ultimate precision limit is given by the Bayesian Cramér-Rao bound (or the van Trees
inequality) \cite{Cramer1946, Rao1995, Van_Trees2004}
\begin{align}
\label{eq:bcrb}
    (\Delta\widetilde{\omega})^2 &\geq [\bar I_C + I_\mathcal{P}]^{-1},
\end{align}
where $\bar I_C \coloneqq \int d\omega \, \mathcal{P}(\omega) I_C(\omega)$, $I_C= \int dx \, p(x|\omega) \left[ \partial_\omega \log p(x|\omega)\right]^2$ is the classical Fisher information (CFI), and $I_\mathcal{P}= \int d\omega \, \mathcal{P}(\omega) \left[ \partial_\omega  \log \mathcal{P}(\omega)\right]^2$ denotes the information from the prior distribution. Since we are interested in the large bandwidth limit, i.e. $\delta\omega\,T \gg 1$, we have that $I_C \gg I_\mathcal{P}$, and Eq. (\ref{eq:bcrb}) reduces to the Cramér-Rao bound (CRB): $(\Delta\widetilde{\omega})^2 \gtrapprox \bar I_C^{-1}$.
Furthermore, from the quantum Cramér-Rao bound (QCRB) \cite{QCRB_Caves}, the CFI is upper bounded by the quantum Fisher information (QFI), i.e. $I_C \leq I_Q$. Then, the QFI is the ultimate limit of precision given an input state, and the benchmark for our parameter estimation. For an $N$ atom CSS, the QFI about $\omega$ is given by $I_Q = 4NT^{2}$, where $N$ is the number of qubits, and $T$ is the interrogation time.
This precision limit is referred to as the Heisenberg limit (HL) with respect to the interrogation time $T$. In contrast, the CFI achieved by repeating a Ramsey protocol for $M=T/\tau$ times is found as $4N \tau^2 \cdot M = 4NT\tau$, which fails to achieve the HL.


We now turn to quantifying the performance of the proposed protocols. We first compute the CFI of this weak measurement scheme. Even though this is a local quantity, it provides a bound on the BMSE through the Bayesian Cramér-Rao bound (Eq. (\ref{eq:bcrb})) and can be calculated analytically. Therefore, we first show that the CFI can approach the QFI. We then demonstrate numerically the saturation of the CFI in the relevant Bayesian setting. 

For this purpose, it is useful to define the parameter $\eta = g^2 T/\tau$, which quantifies the amount of back action on the sensing qubits due to the weak measurements: the back action decay time was found as $\gamma^{-1}\approx\tau/g^2$, therefore, $\eta = T/\gamma^{-1}$ describes how the total interrogation time compares to the decay time.

\begin{figure}
    \centering
\includegraphics[width=\columnwidth]{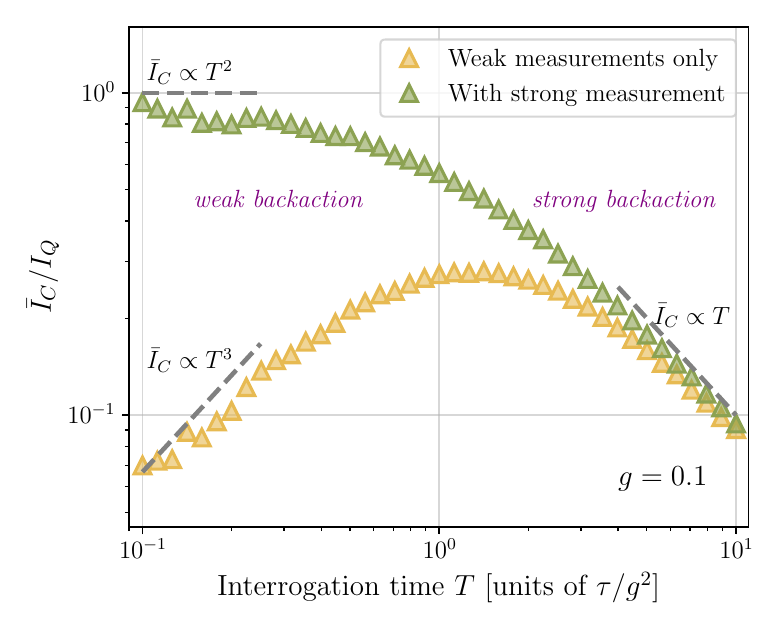}
    \caption{Average classical Fisher information (CFI) $\bar I_C$, normalized by the QFI, $I_Q=4NT^2,$ for the weak-only protocol (plotted in yellow), and weak-with-strong protocol (plotted in green). We fix the measurement strength as $g=0.1$, measurement period as $\tau = 0.1$ s, and vary the interrogation time $T \in [1, 100]$ s.
    In the weak back action regime, i.e. $g^2T/\tau \ll 1$, the CFI for the weak-with-strong protocol saturates $I_Q.$ The CFI for the weak-only protocols is much smaller, yet it increases rapidly as $T^{3}$.
    Eventually, in the strong back action regime, $g^2T/\tau \gg 1$, the two protocols obtain the same CFI of $4NT\tau/g^2$.}
\label{fig:fisher_information_main}
\end{figure}

Let us start with the weak-only protocol, and denote the CFI with this scheme as $\bar I_C^{\text{w}}\left(g,T,\tau \right)$ \footnote{We observed numerically that the CFI is $\omega$-independent in the regime of interest, $\omega \, T \gg  1$, $\omega \tau < \pi/2$}. 
Fig. \ref{fig:fisher_information_main} presents $ \bar I_C^{\text{w}}$ as a function of the interrogation time $T$, for a fixed measurement strength $g$ and period $\tau$. 
We observe a trade-off relation in the behavior of $ \bar I_C^{\text{w}}$: in the weak back action regime,  i.e. for $\eta \ll 1$, $ \bar I_C^{\text{w}}$ can be approximated as $\bar I_C^{\text{w}} \approx \frac{8}{3}Ng^{2}\frac{T^{3}}{\tau}.$
This $T^3$ scaling of the CFI is typical for frequency estimation of classical signal \cite{rife_single_1974, steinhardt_thresholds_1985, knockaert_barankin_1997, schmitt_submillihertz_2017, schmitt_optimal_2021, gardner2025bayesianfrequencyestimationfundamental}. Here, even though the scaling with $T$ is more favorable than HS, the absolute performance relative to the HL is poor since $\eta \ll1$.
Conversely, in the strong back action regime where $\eta \gg 1$, 
the growth of $ \bar I_C^{\text{w}}$ is much more restrained, ultimately converging to $\bar I_C^{\text{w}}\approx4NT/\gamma\approx4N\frac{\tau}{g^{2}}T$. This linear scaling of the CFI is ubiquitous in sequential measurement schemes \cite{gammelmark_fisher_2014,gefen_quantum_2018,yang2023extractable} and it is attributed to the state memory loss induced by the sequential measurements, i.e. the phase becomes uncorrelated with its initial value. We can interpret the overall behavior of the CFI as follows: in the weak back action regime not enough information is extracted from the state, while in the strong back action regime the decoherence induced by the measurements degrades the CFI and prevents achieving HS with respect to time.
The optimal $g$ is thus at the crossover between the two regimes, where the tradeoff between extracted information and dissipation is optimal, and HS can be achieved (see Fig. 4 in \cite{supp}). 


The usual experimental scenario is a fixed total time, $T$, while the weak measurement strength $g$ is tunable. Hence, for a given $T,$ we find that the optimal $g$ corresponds to $g^2T/\tau\approx\sqrt{3/2}$ \footnote{In the relevant regime of $\omega T \gg 1$ and $\omega \tau < \pi/2$.}. The resulting CFI is then $\bar I_C^{\text{w}}\approx 1.11NT^2.$ Hence, we can obtain HS with time using this weak-only scheme with an optimal $g$, where the CFI loses a factor
of $\approx 3.60$ compared to the QFI limit. 

\begin{figure*}
    \centering
\includegraphics[width=\linewidth]{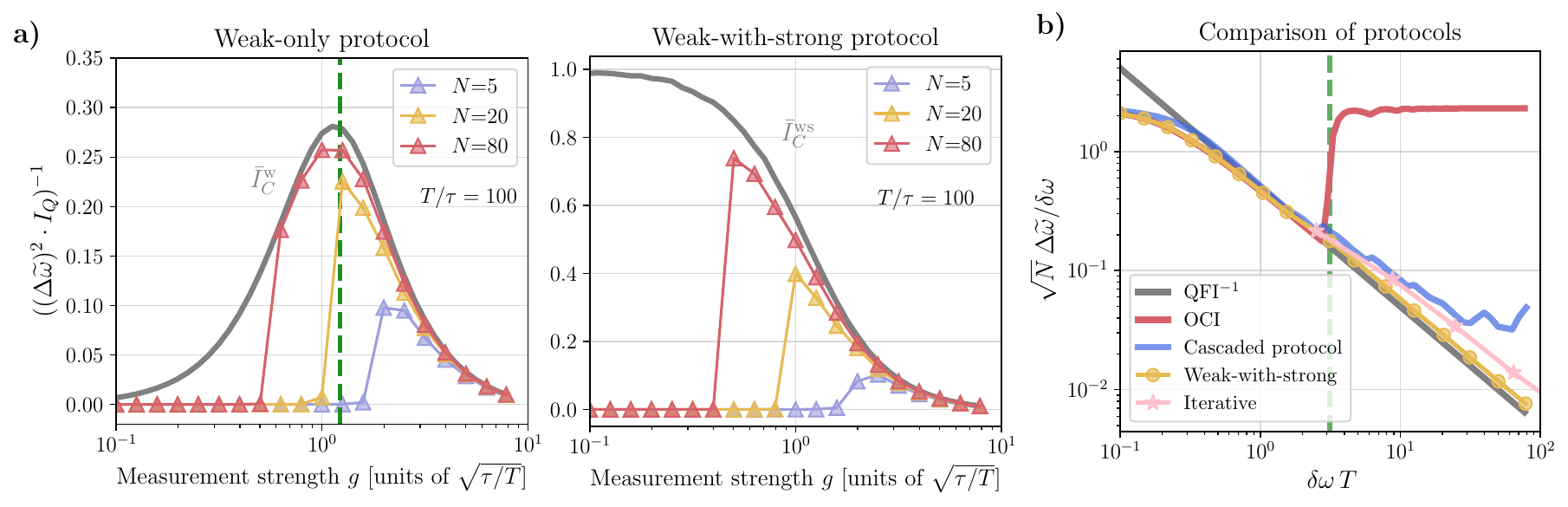}
    \caption{Performance of the proposed protocol with respect to the interrogation time $T$ and the threshold effect. The frequency $\omega$ is sampled from a uniform distribution in $[0, \delta \omega]$, which determines the measurement period as $\tau \lesssim \pi/2\delta\omega$. a) We fix $\delta\omega = 5\pi$ (i.e. $\tau = 0.1$), $T=10$, and plot the inverse of BMSE $(\Delta\widetilde{\omega})^2$ scaled by the QFI $I_Q$ for the two weak measurement protocols. $(\Delta\widetilde{\omega})^2 \cdot I_Q = 1$ signifies saturating the QFI, and $\bar I_C/I_Q$ is plotted with gray, plain lines. We observe a threshold effect such that the attainability of the CFI $\bar I_C^{\text{w}}, \bar I_C^{\text{ws}}$ requires large enough SNR.
    The green, dashed, vertical line corresponds to $T=\sqrt{3/2}\tau/g^2,$ which is a transition point between the weak and strong back action regimes. Achieving HS requires the threshold to occur before this line (which requires $N > 20$ qubits for this case). b) We fix $N=64$, $\delta\omega = \pi$ ($\tau = 0.5$ s) and plot the square-root of the BMSE $\Delta\widetilde{\omega}$ normalized by the prior width $\delta\omega$, and scaled by the square-root of the number of qubits $N$. The vertical green line corresponds to $\delta\omega T = \pi$: for larger interrogation times, phase slips degrade the sensitivity of standard protocols. We compare the performance of: the OCI \cite{macieszczak_bayesian_2014, supp}, the proposed weak-with-strong protocol, the iterative protocol, and the cascaded protocol \cite{rosenband2013exponential}. The latter two demonstrate an extended dynamic range, however the proposed protocol outperforms the cascaded protocol and achieves HS, with an overhead of 1.18 in $\Delta\widetilde{\omega}$. The measurement strength $g$ is optimized for each $T$ for the proposed protocol.}
\label{fig:threshold_mle_victory}
\end{figure*}

The CFI can be improved if, in addition to the weak measurements, a projective (strong) measurement is applied at the end of the interrogation, which is the weak-with-strong protocol.
We denote CFI of this scheme as $\bar I_C^{\text{ws}}\left(g,T,\tau \right)$.
In the weak back action regime ($\eta \ll 1$), $\bar I_C^{\text{ws}} \approx I_Q = 4NT^2$, where it is maximal for $g= 0$, which saturates the QFI limit
(see Fig. \ref{fig:fisher_information_main}).
However, in the strong back action regime ($\eta \gg 1$), the final projective measurement does not add any new information
and we obtain the same linear CFI scaling $\bar I_C^{\text{ws}}\approx 4 NT\tau/g^2$, as we had for $\bar I_C^{\text{w}}$.
Hence, weak measurements only reduce $I_C^{\text{ws}}$,
but as we will show in the following, they are needed to saturate the CFI given a large dynamic range.
A further analysis of the CFI with imperfect weak measurements can be found in \cite{supp}.

{\textit{Saturability of the CRB and the threshold effect---}}So far we have shown that the CFI of both of the weak measurement protocols, $\bar I_{C}^{\text{w}}, \bar I_C^{\text{ws}}$, yield HS with appropriate values of
$g,\tau, T$. However, these quantities
correspond to local estimation of $\omega,$
and they are not necessarily saturable in the relevant large bandwidth limit, $\delta\omega T \gg 1$.
Hence, we study the minimal $\eta,N$ for which $\bar I_C^{\text{w}},\bar I_C^{\text{ws}}$ are saturated,
given a uniform prior distribution in $[0, \delta\omega]$. In particular, we focus on whether they can be saturated for small enough $\eta$ such that HS is achieved. Note that the prior distribution implicitly determines the measurement period for the weak measurement protocols, with $\tau \lesssim \pi/2\delta\omega$. 


A similar problem was studied extensively in  classical sensing, where a signal $s(t_n)=A \cos(2 \omega t_n)+\nu_n$
is sampled at times $\{t_n=n \tau\}_{n=1}^{M}$
\cite{rife_single_1974,steinhardt_thresholds_1985}.
Here, $A$ is the amplitude of the signal and $\{\nu_n\}_{n=1}^{M}$
are i.i.d. Gaussian  random variables, $\nu_{n}\sim N\left(0,\sigma^{2}\right)$.
It was shown that this classical frequency estimation problem is plagued by a threshold effect with respect to the signal to noise ratio (SNR), defined as $\frac{A^{2}}{\sigma^{2}}\frac{T}{\tau}.$
For low SNR,
the posterior distribution is very close to the prior distribution, and the MSE is thus very close the prior variance. 
Above a certain threshold of the SNR, there is a sharp drop in the MSE, and it converges to the CRB. This is a well-known effect in signal estimation, studied in e.g. Refs. \cite{rife_single_1974, steinhardt_thresholds_1985, knockaert_barankin_1997}. 

In our case we also sample a stochastic, time-dependent signal, through the weak measurements. The weak back action regime in our setting is equivalent to sampling a classical signal due to the negligible back action.
We thus expect a similar threshold behavior, where the SNR in the limit of $g \ll1$ can be defined as $Ng^{2}T/\tau = N \eta.$
This SNR threshold effect is observed numerically, and it is illustrated in Fig. \ref{fig:threshold_mle_victory}a (see also Fig. 6 in \cite{supp}). We plot the
BMSE $(\Delta\widetilde{\omega})^2$ obtained with the maximum likelihood estimator (MLE) as a function of $T$ for varying numbers of qubits $N$ \footnote{We use the optimal Bayesian estimator \cite{kaubruegger_quantum_2021, direkci2024heisenberg} for $\delta\omega T < \pi$, since the prior information is significant in this regime.}.
$(\Delta\widetilde{\omega})^{-2}$ is compared to the CFI
as a benchmark, plotted with gray lines.
As the SNR is increased, 
e.g. by increasing $T$ as in Fig. ~\ref{fig:threshold_mle_victory}a, 
the BMSE starts to rapidly decrease and saturates the CRB.

In order to approach $I_Q$ and achieve HS with time, the CRB needs to be saturated in the regime where $\bar I_C^{\text{ws}} \approx I_Q$. Therefore, reaching this limit requires (i) weak measurement back action, $\eta < 1,$ and (ii) large enough SNR, $N \eta >1$. As shown in Fig. \ref{fig:threshold_mle_victory}a, these two conditions are incompatible for small $N$: the saturability occurs only in the strong back action regime. However, there exists a minimal $N$ after which these two conditions are satisfied and HS can be achieved. 
In particular, we observe that as $N$ 
increases, the saturability occurs at smaller $\eta$.

Let us analytically find the minimal $N$ for which the CRB can be saturated. We aim to compute the parameters of the weak-with-strong protocol for which the BMSE is close to the CFI, i.e. $\left(\Delta\widetilde{\omega}\right)^{-2}=\left(1-\epsilon\right)\bar I_{C},$ $\epsilon \ll 1$.
We work in the limit of $\eta \ll 1$ and a large number of probes, $N \gg 1$, for which the problem can be approximated as the sampling of a classical signal. 
We find that $\left(\Delta\widetilde{\omega}\right)^{-2}=\left(1-\epsilon\right)\bar I_{C}$
is attained for
\begin{align}
\label{eq:threshold_cond_mt}
    N\eta \approx 2\ln{\left(\frac{\pi^{3/2}}{48\epsilon} \right)} + 6\ln{\left(\frac{T}{\tau}\right)}.
\end{align}
For any system parameters $g, T, \tau,$
this expression provides 
the minimal number of probes $N$ for which we are $\epsilon$-close to the 
CFI. 
Since we can always tune $g$ such that $\eta$ remains constant,
then the minimal $N$, for any given $\epsilon$, grows as $6\ln\left[\left(2\delta\omega T\right)/\pi\right]$.
Hence it grows logarithmically with the number of phase wraps. While this logarithmic dependence is in line with the scaling of $N$ in previous schemes \cite{kessler_heisenberg-limited_2014},
we obtain a smaller overhead.

Finally, we want to explicitly show that this protocol achieves an almost optimal BMSE for long times, i.e. $\delta\omega T \gg1$, and that it outperforms existing methods.
To this end, we plot in Fig. \ref{fig:threshold_mle_victory}b the square-root of the BMSE $\Delta \widetilde{\omega},$
as a function of interrogation time for different sensing schemes for $N=64$.
Our weak-with-strong protocol is compared to the QFI bound, to the
OCI, and to two more conventional dynamic range extension schemes: the iterative and the cascaded protocol \cite{rosenband2013exponential, kessler_heisenberg-limited_2014, shaw_multi-ensemble_2023, direkci2024heisenberg}. The iterative protocol scales the interrogation time by a constant factor $\nu>1$ iteratively, whereas the cascaded protocol employs groups of atoms with shorter interrogation times $T/2^i, \, i>0$, 
which are used to 
prevent phase slips \cite{supp}. In the Figure, we optimize over $\nu$, which is set to 2.51.

We first observe that our weak-with-strong protocol performs very close to the QFI bound for the entire range of interrogation times,
with a maximal overhead of $1.18$ in $\Delta\widetilde{\omega}$ (compared to 1.75 of the iterative scheme).
We thus obtain an almost optimal HS that is immune to phase slip errors and persists in the limit of large $\delta \omega T.$
This is in sharp contrast to the OCI that suffers from phase slips errors and becomes inefficient for $\delta \omega T>\pi.$
We also observe that our protocol considerably outperforms the cascaded protocol.
The intuitive reason behind this metrological gain is that, while the cascaded protocol allows optimization over only a finite set of partitions into blocks, the proposed weak measurement scheme offers greater tunability, as the optimization is over the continuous weak measurement strength $g$ \footnote{Note that if we plot $\Delta\tilde{\omega}$ for larger $T$, the optimal $g$ will fall into the strong backation regime, causing the transition from HS to a $T$ scaling in the obtained precision.}.


\textit{Conclusions and outlook---}In this Letter, we developed and analyzed a weak measurement Ramsey protocol that achieves optimal precision over an extended dynamic range asymptotically. By optimizing over the weak measurement strength, we were able to achieve a nearly optimal sensitivity and HS in the large bandwidth limit, $\delta\omega T \gg 1$, given a large enough number of probes. 
We identified two separate conditions for obtaining a high sensitivity and a large bandwidth, and demonstrated that they can be satisfied simultaneously. 

Some open questions and directions that can potentially improve the proposed protocol are as follows:
first, we can reduce the number of ancillas
by coupling only a subset of sensors to ancillas
or coupling several sensors to the same ancilla.
We can also improve the BMSE in the low SNR limit by adding more ancillas as memory qubits and performing a collective measurement on them \cite{yang2023efficient, allen2025quantum, gardner2025bayesianfrequencyestimationfundamental}. This may allow achieving HS for smaller $N.$ Finally, we can mitigate the effect of imperfect measurements by using error correction \cite{Len2022,Zhou_noisy_meas,carmel2023hybrid, ouyang2024robust}. Weak measurements could also be combined with entangled states \cite{borregaard_near-heisenberg-limited_2013} to achieve HS both with time, and the number of qubits, for an extended bandwidth.

\textit{Acknowledgements---}We acknowledge helpful discussions with Yanbei Chen, Ran Finkelstein, James W. Gardner, Lee P. McCuller, Alex Retzker, Nelson Darkwah Oppong, and Denis V. Vasilyev.
TG acknowledges funding from the quantum science and technology early-career grant of the Israeli council for higher education.
We acknowledge funding from the Army Research Office MURI program (W911NF2010136), TINA QC (W911NF2410388), and the Institute for Quantum Information and Matter, and the NSF Physics Frontiers Center (NSF Grant PHY-1733907).


\clearpage
\onecolumngrid
\begin{center}
\textbf{\large Supplemental Material}
\end{center}
\setcounter{equation}{0}
\setcounter{figure}{0}
\setcounter{table}{0}
\setcounter{page}{1}
\makeatletter
\renewcommand{\theequation}{S\arabic{equation}}
\renewcommand{\thefigure}{S\arabic{figure}}

\tableofcontents

\section{Optimal classical interferometry and phase slips (Fig. 1c)}
\label{app:oci}
In this appendix, we explain how to calculate the optimal classical interferometer (OCI).
OCI refers to the optimal BMSE given a coherent spin state (after the free evolution), $|+,\omega\rangle =2^{-N/2}\left(|0\rangle+|1\rangle e^{-2i\omega T}\right)^{N}$, and a frequency prior distribution, $\mathcal{P}_{\delta\omega}(\omega),$
when optimized over all possible measurements.
The derivation of this bound is based on Ref. \cite{macieszczak_bayesian_2014}, which showed how to calculate the optimal BMSE for any input state $\rho_{\omega}$ and prior $\mathcal{P}_{\delta\omega} (\omega).$
For completeness, we provide here a derivation. Given a density matrix $\rho_\omega$ and assuming a POVM of $\left\{ \Pi_{x}\right\} _{x},$ the BMSE is given by
\begin{align}
\label{supp_eq:bmse_general}
(\Delta\tilde{\omega})^{2}=\int d\omega\;dx\;\mathcal{P}_{\delta\omega}\left(\omega\right)\text{Tr}\left(\rho_{\omega}\Pi_{x}\right)\left(\omega_{\text{est}}\left(x\right)-\omega\right)^{2}.    
\end{align}
This expression can be simplified to 
\begin{align}
\label{supp_eq:bmse_opt_estimator}
(\Delta\tilde{\omega})^{2}=\text{var}\left(\mathcal{P}_{\delta\omega}\left(\omega\right)\right)-\int dx\frac{\text{Tr}\left(\Pi_{x}\bar{\rho}'\right)}{\text{Tr}\left(\Pi_{x}\bar{\rho}\right)}, 
\end{align}
where $\bar{\rho}=\int\mathcal{P}_{\delta\omega}\left(\omega\right)\rho_{\omega}\,d\omega,$
$\bar{\rho}'=\int\omega\mathcal{P}_{\delta\omega}\left(\omega\right)\rho_{\omega}\,d\omega.$
This simplification assumes $\int\omega\mathcal{P}_{\delta\omega}\left(\omega\right)=0,$ and this can be always satisfied when the encoding is unitary by shifting the parameter by another unitary at the end of the interrogation.
The bound now follows from the observation that
\begin{align}
\label{supp_eq:estimator_upper_bound}
\int dx\frac{\text{Tr}\left(\Pi_{x}\bar{\rho}'\right)}{\text{Tr}\left(\Pi_{x}\bar{\rho}\right)}\leq\text{Tr}\left(L^{2}\bar{\rho}\right),    
\end{align}
where $1/2\left\{ L,\bar{\rho}\right\} =\bar{\rho}'.$
This bound is tight, and it is saturated by taking $\left\{ \Pi_{x}\right\} _{x}$ to be projections onto eigenspaces of $L$.
We can thus conclude that the BMSE optimized over all possible measurements is given by:
\begin{align}
(\Delta\tilde{\omega})^{2}_{\text {min}}=\text{var}\left(\mathcal{P}_{\delta\omega}\left(\omega\right)\right)-\text{Tr}\left(L^{2}\bar{\rho}\right).    \label{eq:OCI_bound_general}
\end{align}
It is straightforward to apply this bound to the coherent spin state: taking $\rho_{\omega}=|+,\omega\rangle\langle+,\omega|,$ we numerically calculate the relevant $L,$ and insert it in in Eq. (\ref{eq:OCI_bound_general}) to obtain the OCI.

In the limit of $\delta\omega T \gg 1$, where $\delta\omega$ is the prior width, we observe from Fig. 1c in the main text that the BMSE $(\Delta\widetilde{\omega})^2$ increases significantly, and converges to a constant value. In this limit, no information is obtained from the frequency estimation due to phase slips, as the phase encoding unitary is periodic in $\omega T$. Let us demonstrate the effect of phase slips mathematically, using \cref{supp_eq:bmse_general,supp_eq:bmse_opt_estimator,supp_eq:estimator_upper_bound,eq:OCI_bound_general}. We assume a uniform prior distribution in $[-\delta\omega/2, \delta\omega/2]$, such that $\text{var}\left(\mathcal{P}_{\delta\omega}\left(\omega\right)\right) = (\delta\omega)^2/12$. We first observe that for this uniform prior, $\bar \rho_{kl} \propto \text{sinc}[\delta\omega T (k-l)]$, and $\bar\rho'_{kl} \propto [\delta\omega (k-l)\cos{(\delta\omega(k-l))} - \sin{(\delta\omega(k-l))}]/[\delta\omega T(k-l)]^2$, $k, l \in \mathbb{N}$. Then, in the limit of $\delta\omega T \gg 1$, $\bar \rho$ will approximately be diagonal, and $\bar \rho'$, $L$ will approach the null operator.
We will therefore have from Eq. (\ref{eq:OCI_bound_general}) that $(\Delta\tilde{\omega})^{2}_{\text {min}}\approx(\delta\omega)^2/12$, which results in the behavior in Fig. 1c in the main text.


\section{Realization of a Weak Measurement}
\label{app:realization}

\begin{figure}
    \centering
    \begin{minipage}{0.5\textwidth}
    \centering
\begin{quantikz}
\lstick{$|+\rangle$}  &  \qw \gategroup[2,steps=5,style={inner
sep=6pt, dashed, fill=yellow!15}, background]{weak measurement} & \qw  & \gate{H} & \ctrl{1} & \gate{H} & \meter{}& \\[-3mm]
& &  \lstick{$|\uparrow_{y}\rangle$}& \gate{R_x(-g)}&\gate{R_x(2g)} & \meter{} & &
\end{quantikz}   
\end{minipage}
\hspace{-0.5cm}
\begin{minipage}
{0.5\textwidth}
\centering
\includegraphics[width=0.9 \textwidth]{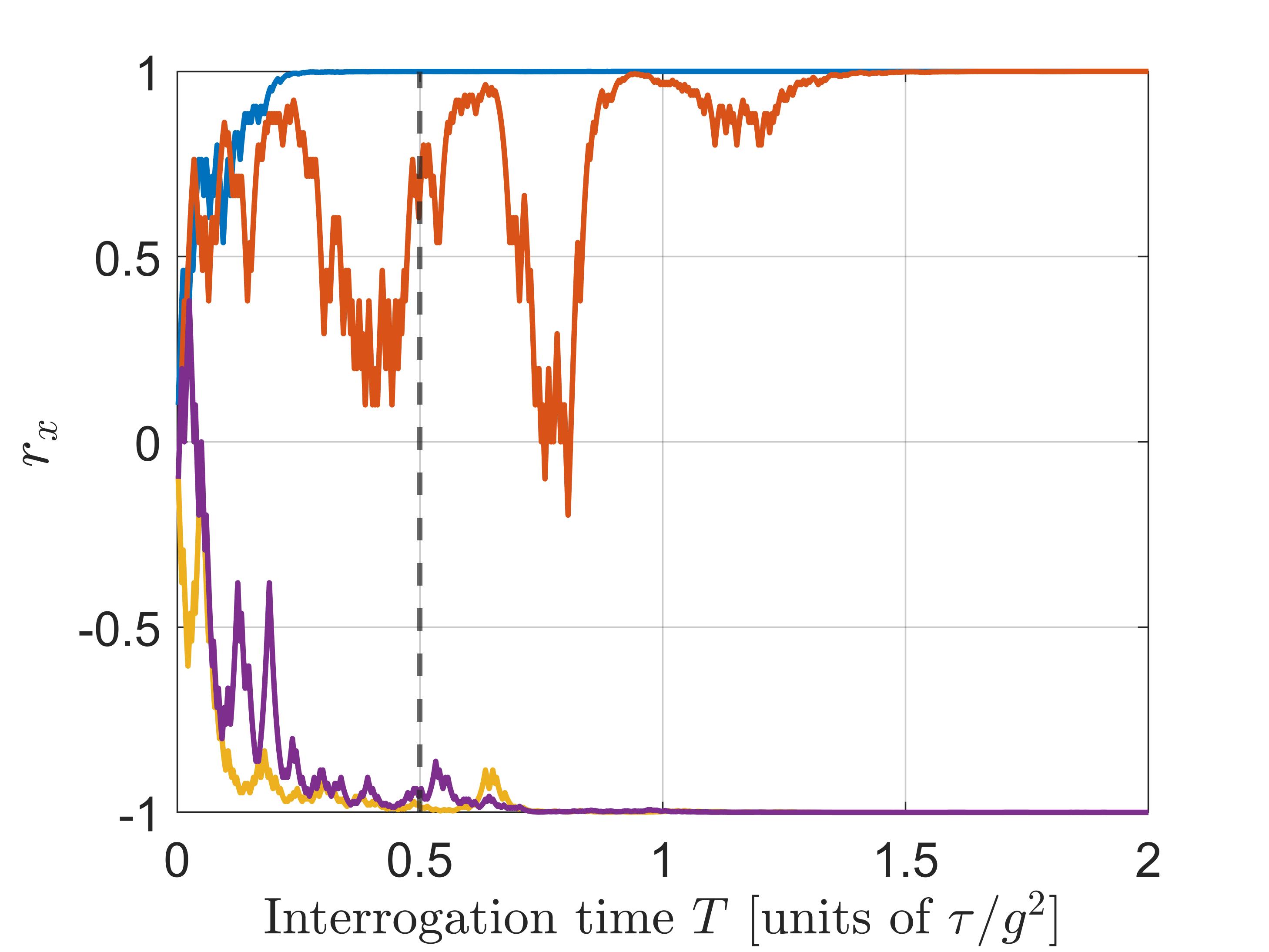}
\end{minipage}
    \caption{Left: Quantum circuit for implementation of a weak Pauli $\sigma_{x}$ measurement.
    This circuit consists of a conditional rotation
    of the ancilla based on the $\sigma_x$ eigenstate of the probe ($H$ stands for the Hadamard gate). We define the single-qubit rotation about the x-axis as $R_x(g) = \exp(i g \sigma_x)$.  
    Right: $r_x$ as a function of the interrogation time $T$, for different realizations of weak measurements, shown by curves with varying colors. $r_x$ undergoes a biased random walk with absorbers at $r_x=\pm 1.$
    The sequential application of weak measurements converges to a strong measurement of $\sigma_x$ and to a collapse to one of the eigenstates of $\sigma_x$. In this illustration, we set $g=0.05,$ $\tau=0.01$ s and vary $T$ in $[0, 8]$ s.  }\label{fig:weak_meas_circuit}
\end{figure}

We consider a weak Pauli $\sigma_{x}$ measurement of a qubit. 
This measurement is given by the following two Kraus operators
\begin{align}
\label{eq:kraus_ops}
    K_{\pm} = \frac{1}{\sqrt{2}} \left[ \cos{(g)} I \pm \sin{(g)} \sigma_{x} \right]= \frac{1}{\sqrt{2}} \begin{bmatrix}
        \cos{(g)} & \pm \sin{(g)} \\
        \pm \sin{(g)} & \cos{(g)}
    \end{bmatrix}.
\end{align}
Here, $g$ is a unitless variable denoting the strength of the measurement: $0\leq g \leq \pi/4$ interpolates between weak and strong measurement, where the latter is achieved for $g=\pi/4.$ 
Following a weak measurement, the state of the qubit $\rho$ is updated as
\begin{align}
    \rho \rightarrow \frac{K_{\pm} \, \rho \, K_{\pm}^\dagger}{\text{Tr}[K_{\pm} \, \rho \, K_{\pm}^\dagger]}
\label{eq:update_rule_weak_meas}    
\end{align}
depending on the measurement result. 
The weak measurement described by the Kraus operators in Eq. (\ref{eq:kraus_ops}) can be realized with any digital quantum device using the circuit in Fig. \ref{fig:weak_meas_circuit}. 
This protocol employs an ancilla qubit, 
initialized in the state $\ket{\uparrow_y} = (\ket{1} + i \ket{0})/\sqrt{2}$.
The ancilla is then entangled with the probe (the sensing qubit) using a controlled rotation unitary 
\begin{align}
    U = e^{-i g \, \sigma_x \otimes \sigma_x} = \cos{(g)} I - i \sin{(g)} \, \sigma_x \otimes \sigma_x,
\end{align}
i.e. an $R_{x}(g)$ rotation of the ancilla conditioned on the state of the probe in the $\sigma_{x}$ basis.
The joint state of the probe and the ancilla is then given by $U \rho \otimes \ket{\uparrow_y} \bra{\uparrow_y} U^\dagger$ after the entangling interaction $U$.
Measuring the ancilla 
in the basis $\{\ket{0}, \ket{1}\}$ yields the Kraus operators in Eq. (\ref{eq:kraus_ops}).

Let us first illustrate the fact that this is indeed a weak measurement of $\sigma_{x}$, and find the convergence rate to a strong measurement.
This can be seen from the average dynamics, i.e. the dynamics of the density matrix when averaged over all possible measurement results. It can be described by the following channel:
\begin{align}
\rho\left(T\right)=\Lambda^{T/\tau}\left(\rho\left(0\right)\right),
\end{align}
where $\Lambda\left(\rho\right)=K_{+}\rho K_{+}^{\dagger}+K_{-}\rho K_{-}^{\dagger}$, and $\Lambda^m\left(\rho\right)$ denotes applying the channel $m$ times. The initial state for the qubit can be written as $\rho\left(0\right)=1/2\left(I+\vec{r}\cdot\vec{\sigma}\right),$ where $\vec{r} = (r_x, r_y, r_z)$, and $|\vec{r}| \leq  1$. Then, using the Kraus operators in Eq. (\ref{eq:kraus_ops}), we obtain that $\rho\left(\tau\right)=1/2\left[I+r_{x}\sigma_{x}+\cos\left(2g\right)\left(r_{y}\sigma_{y}+r_{z}\sigma_{z}\right)\right]$. Hence,
\begin{align}
\rho\left(T\right)=1/2\left[I+r_{x}\sigma_{x}+\cos\left(2g\right)^{T/\tau}\left(r_{y}\sigma_{y}+r_{z}\sigma_{z}\right)\right].    
\end{align}
The density matrix thus converges to a completely dephased density matrix (in the Pauli $\sigma_{x}$ basis), which corresponds a $\sigma_{x}$ measurement.
The convergence rate to a strong measurement is thus $\gamma_{\text{meas}}=-\text{log}\left(\cos\left(2g\right)\right)/\tau \approx 2g^2/\tau$ for $g \ll 1$. 

Let us analyze the stochastic evolution of individual trajectories given by the weak measurements outcomes.
In each trajectory, $r_x$ undergoes a random walk with two absorbers at $r_x=\pm1$ that correspond to $|\pm\rangle = \frac{1}{\sqrt{2}}(\ket{1} \pm \ket{0})$, respectively. The random walk is described by
\begin{align}
r_{x}\rightarrow r_{x}\pm\frac{\sin\left(2g\right)\left(1-r_{x}^{2}\right)}{1+r_{x}\sin\left(2g\right)}\;\;\text{with }\;\;p(x|\omega)=\frac{1}{2}\left(1+(-1)^x r_{x}\sin\left(2g\right)\right),
\end{align}
where both the step size, $\frac{\sin\left(2g\right)\left(1-r_{x}^{2}\right)}{1+r_{x}\sin\left(2g\right)}$, and the probabilities, $p(x|\omega),$ depend on $r_x$. The two measurement probabilities are parametrized with $x=0,1$ that correspond to a positive and a negative step, respectively. This stochastic dynamics leads to trapping of $r_x$ in one of the absorbers at $r_x=\pm1,$ which is the collapse of the state to $|\pm\rangle$.
Numerical examples of this process are presented in Fig. \ref{fig:weak_meas_circuit}.

\section{Ramsey interferometry with weak measurements}
To perform Ramsey interferometry with weak measurements,
the qubit is initialized in the state $\ket{+} = (\ket{0} + \ket{1})/\sqrt{2}$. It evolves freely under the Hamiltonian $\mathcal{H}(\omega) = \omega \, \sigma_{z}$ for a duration of $\tau$, where $\sigma_z$ is the Pauli $z$ operator, after which a weak $\sigma_{x}$ measurement is applied. This process repeats for $m=\lfloor T/\tau \rfloor$ times.
It can be therefore described with the following two Kraus operators: $K_{\pm}^{\left(\omega\right)}=K_{\pm}V_{\omega},$ where $V_{\omega}=\cos\left(\omega\tau\right)I-i\sin\left(\omega\tau\right)\sigma_{z}$ is the free evolution unitary, and $K_{\pm}$ are given in Eq. (\ref{eq:kraus_ops}). 

To understand this evolution better, let us study the average dynamics described by the channel $\Lambda_{\omega}^{T/\tau}\left(\rho\left(0\right)\right),$ where 
\begin{align}
\Lambda_{\omega}\left(\rho\right)=K_{+}^{\left(\omega\right)}\rho K_{+}^{\left(\omega\right)\dagger}+K_{-}^{\left(\omega\right)}\rho K_{-}^{\left(\omega\right)\dagger}.
\label{eq:avg_dynamics_channel}
\end{align}
Similar to the previous Section, this channel represents averaging over all possible ancilla measurements during time $0<t < T$. Expanding $\rho\left(0\right)=1/2\left(I+\vec{r}\cdot\vec{\sigma}\right),$
we obtain
\begin{align}
\left(\begin{array}{c}
r_{x}\left(\tau\right)\\
r_{y}\left(\tau\right)\\
r_{z}\left(\tau\right)
\end{array}\right)=\left(\begin{array}{ccc}
\cos\left(2\omega\tau\right) & \sin\left(2\omega\tau\right) & 0\\
-\cos\left(2g\right)\sin\left(2\omega\tau\right) & \cos\left(2g\right)\cos\left(2\omega\tau\right) & 0\\
0 & 0 & \cos\left(2g\right)
\end{array}\right)\left(\begin{array}{c}
r_{x}\\
r_{y}\\
r_{z}
\end{array}\right).    
\end{align}
The dynamics of $r_z$ is decoupled from the that of $r_x,r_y,$ and since $r_z(0)=0$, it remains $0.$
Restricting ourselves to the block matrix of $r_x,r_y,$
its eigenvalues are 
\begin{align}
\label{eq:avg_dynamics_eigenvals}
\cos\left(2g\right)^{1/2}e^{\pm i\alpha} \; \text{ with 
 } \alpha = \arccos{ \left( \frac{\cos{(g)}^2 \cos{(2 \omega \tau)}}{\sqrt{\cos{(2g)}}} \right)}.     
\end{align}
This implies that $r_x, r_y$ decay with a rate of $\gamma=-\frac{1}{2\tau}\log\left(\cos\left(2g\right)\right)$
and oscillate with a frequency of $\alpha/\tau.$ Let us denote the probability of ancilla measurements at time $t = k \tau$, $k \in \mathbb{N}^+$ as $p(x_k |\omega)$, where $x_k = 0, 1$ correspond to the Kraus operators $K^{(w)}_+$ and $K^{(w)}_-$, respectively.
The full solution of $r_x(k\tau)$ and $p(x_k  = 0|\omega)$ is then given by
\begin{align}
\begin{split}
&r_{x}\left(k\tau\right)=\cos\left(2g\right)^{k/2}A\cos\left(\alpha t+\phi\right),\\
&p(x_k  = 0|\omega)=\frac{1}{2}\left(1+\sin\left(2g\right)r_{x}\left(k\tau\right)\right)=\frac{1}{2}\left[1+\sin\left(2g\right)\cos\left(2g\right)^{k/2}A\cos\left(\alpha k+\phi\right)\right]
\end{split}
\label{eq:averaged_dynamics}
\end{align}
where
\begin{align}
A=\sqrt{\frac{\cos{(2g)} \sin{(2\omega \tau)}^2}{\cos\left(2g\right)-\cos\left(g\right)^{4}\cos\left(2\omega\tau\right)^{2}}}, \;\; \phi =\arcsin\left(\frac{-\sin{(g)}^{2}\cos{(2\omega\tau)}}{\sqrt{\cos{(2g)}\sin{(2\omega\tau)}^2}}\right).
\end{align}
Here, $k$ denotes the number of the measurement, and $\alpha$ is given in Eq. (\ref{eq:avg_dynamics_eigenvals}).
For $g^2T/\tau \ll 1$, the parameters $\alpha,\,A,\,\phi$ can be approximated as $\alpha  = 2\omega\tau + O(g^4), \; A = 1 + O(g^4),$ and $\phi = O(g^2)$. Then, $p(x_k  = 0|\omega) \approx [1+\sin\left(2g\right)\cos\left(2g\right)^{k/2}\cos\left(2\omega k\tau\right) ]/2$.

Let us now study the stochastic qubit trajectories defined by a string of measurement results. Contrary to the average dynamics, we keep track of the ancilla measurements, which indicate the trajectory of the sensing qubit.
For this case, we can show that the state of the qubit and the measurement probabilities can be calculated iteratively.
We first prove the following simple claim:
The stochastic process of $K_{\pm}^{(\omega)}$ preserves the state in the $\sigma_{x}$-$\sigma_{y}$ plane; i.e. if the initial state is in this plane, it will remain there at all time (ignoring decoherence effects, e.g. amplitude damping).

\begin{figure}
    \centering
\includegraphics[width=0.5\linewidth]{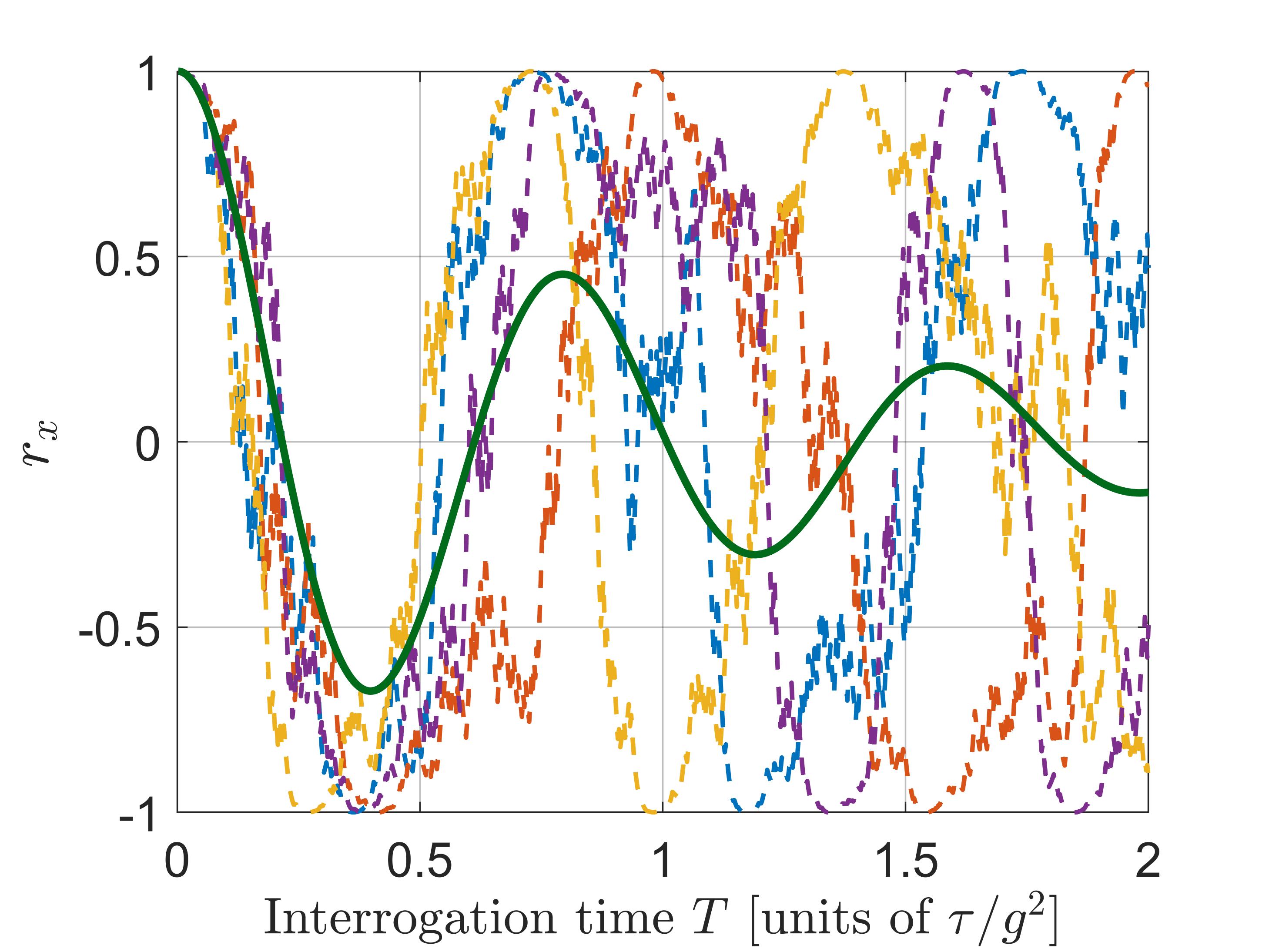}
    \caption{$r_x$ as a function of time for: different trajectories (dashed lines), and averaged dynamics (solid green line) of Ramsey with weak measurements. The weak measurements induce a random walk of the phase that manifests as dephasing of the averaged dynamics. Because of this dephasing, $r_x \rightarrow 0$ in the limit of infinitely many measurements, signifying that we lose the qubit contrast completely. In this illustration, the measurement strength is $g =0.05$, the measurement period is $\tau = 0.01$ s, and the we vary the interrogation time in $T\in[0, 8]$ s.}
    \label{fig:trajectories_vs_averaged}
\end{figure}

To show this we use the update rule of Eq. (\ref{eq:update_rule_weak_meas}), which in this case reads $\rho\mapsto K_{\pm}^{\left(\omega\right)}\rho K_{\pm}^{\left(\omega\right)\dagger}/\text{Tr}(K_{\pm}^{\left(\omega\right)}\rho K_{\pm}^{\left(\omega\right)\dagger}).$ Hence, it suffices to show that the (unnormalized) operation $K_{\pm}^{\left(\omega\right)}\rho K_{\pm}^{\left(\omega\right)\dagger}$
does not take the state out of the $\sigma_{x}$-$\sigma_{y}$ plane. Assuming the qubit to be in the state $\rho=1/2\left(I+\vec{r}\cdot\vec{\sigma}\right),$ where $\vec{r} = (r_x, r_y, r_z)$, and $|\vec{r}| \leq  1$, this operation induces the following mapping of $\vec{r}$:
\begin{subequations}
\begin{align}
&r_{x}\mapsto\cos\left(2\omega\tau\right)r_{x}+\sin\left(2\omega\tau\right)r_{y}\pm\sin\left(2g\right)\\
&r_{y}\mapsto-\cos\left(2g\right)\sin\left(2\omega\tau\right)r_{x}+\cos\left(2g\right)\cos\left(2\omega\tau\right)r_{y}\\
&r_{z}\mapsto\cos\left(2g\right)r_{z}
\label{supp_eqn:update_bloch_vector}
\end{align}
\end{subequations}
since $r_{z}\mapsto r_{z}\cos\left(2g\right)$, a state initialized in the $\sigma_{x}-\sigma_{y}$ plane ($r_{z}=0$) will remain there. Hence, given an initial state of $|+\rangle$, the state of the qubit remains in the $\sigma_{x}$-$\sigma_{y}$ plane for all possible measurement results, at all time. Since the state also remains pure, it is therefore fully characterized by its angle in the plane, $\theta$, and it can be written as  $\rho=\frac{1}{2}\left(I+\cos\left(\theta\right)\sigma_{x}+\sin\left(\theta\right)\sigma_{y}\right).$ For a given trajectory, let us denote as $\rho_{k}$ and $\theta_k$ the state and the angle before the $k$-th measurement.
The weak measurement probabilities at the $k$-th stage are then
given by
\begin{align}
\label{supp_eqn:prob_weak_meas}
p(x_{k}|\omega):=\text{Tr}\left(K_{(-1)^{x_{k}}}\rho_{k}K_{(-1)^{x_{k}}}^{\dagger}\right)=\frac{1}{2}\left[1+(-1)^{x_{k}}\sin{(2g)}\cos{(\theta_{k})}\right]
\end{align}
where $x_{k} = 0, 1$, $k \in \mathbb{N}^+$. Given that the $k$-th measurement outcome is $x_k$, the state is updated as follows:
\begin{align}
\rho_{k+1}=\frac{1}{p\left(x_{k}|\omega\right)}V_{\omega}K_{(-1)^{x_{k}}}\rho_{k}K_{(-1)^{x_{k}}}^{\dagger}V_{\omega}^{\dagger}.
\end{align}
Inserting $\rho_{k}=\frac{1}{2}\left(I+\cos\left(\theta_{k}\right)\sigma_{x}+\sin\left(\theta_{k}\right)\sigma_{y}\right),$
we obtain the
following recursive relation for $\theta_{k}$:
\begin{align}
\theta_{k+1}=\text{arg}\left(\cos\left(\theta_{k}\right)+(-1)^{x_{k}}\sin\left(2g\right)+i\cos\left(2g\right)\sin\left(\theta_{k}\right)\right)-2\omega\tau.
\label{eq:phase_update}
\end{align}
Note that in the limit of $g\ll1$, this relation reduces to $\theta_{k+1}=\theta_{k}-(-1)^{x_{k}}2g\sin\left(\theta_{k}\right)-2\omega\tau+O\left(g^{2}\right),$
hence $\theta_{k}$ undergoes a scaled random walk, with a step size of $2g\sin\left(\theta_{k}\right)$.
The angle and measurement probabilities can be thus calculated iteratively using Eq. (\ref{eq:phase_update}). In Fig. \ref{fig:trajectories_vs_averaged}, we sample different realizations of $r_x$ for a single qubit, and plot them as a function of the interrogation time $T$ with dashed lines. For the numerical simulations, we use the parameters of $g = 0.05$, $\tau =0.01 $ s, and $T \in [0, 8]$ s. We also plot $r_x$ for the averaged dynamics, given in Eq. (\ref{eq:averaged_dynamics}). We observe that $r_x$ oscillates between $\pm 1$ for the individual qubit trajectories, whereas the $r_x$ for the averaged dynamics decays with time, signifying the existence of dephasing due to the weak measurements.


\section{Cramer-Rao bounds}

\subsection{Fisher Information}

Given an $\omega$ dependent distribution $\left\{ p\left(x|\omega\right)\right\} _{x},$ the Cramér-Rao bound provides a fundamental precision limit in estimating $\omega$ by sampling $\left\{ p\left(x|\omega\right)\right\} _{x}$. It states that the MSE of any unbiased estimator satisfies: $\left(\Delta\omega\right)^{2}\geq I_C^{-1},$
where $I_C$ is the classical Fisher information (CFI) about $\omega$ given by \cite{Cramer1946, kay1993fundamentals}:
\begin{align}
\label{eq:fi_defn}
    I_C = \text{E}\left[\left(\frac{\partial}{\partial \omega}\text{ln} p(x|\omega) \right)^2 \right] = \sum_{x} \frac{1}{p(x|\omega)} \left( \frac{\partial p(x|\omega)}{\partial \omega} \right)^2.
\end{align}
This bound is, however, not necessarily tight. It holds only if the map $\omega\mapsto p(x|\omega)$ is injective in the domain of $\left[\omega_{\text{min}},\omega_{\text{max}}\right].$
Given that this map is injective, this bound is attainable for a large number of samples of $p(x|\omega)$ with maximum likelihood estimation \cite{kay1993fundamentals}. Given the CFI about the parameter $\omega$, $I_C$, and a prior distribution, we can write down the average CFI as
\begin{align}
\label{eq:avg_cfi_defn}
    \bar I_C \coloneqq \int d\omega \, \mathcal{P}(\omega) I_C(\omega),
\end{align}
which bounds the Bayesian mean squared error (BMSE) of the estimation as $(\Delta\widetilde{\omega})^2 \geq [\bar I_C + I_\mathcal{P}]^{-1}$, where $I_\mathcal{P}= \int d\omega \, \mathcal{P}(\omega) \left[ \partial_\omega  \log \mathcal{P}(\omega)\right]^2$, as given in Eq. (2) in the main text. In quantum parameter estimation problems, the classical distribution is replaced by an $\omega$ dependent quantum state $\rho(\omega).$ The MSE is then lower bounded by the quantum Fisher information (QFI):
$\left(\Delta\omega\right)^{2}\geq I_{Q}^{-1},$ where $I_Q$ is the QFI. The QFI for a general $\rho(\omega)$ is given by \cite{QCRB_Caves}
\begin{align}
    I_{Q}=\sum_{j,k}\frac{2}{p_{j}+p_{k}}|\langle j|\partial_{\omega}\rho|k\rangle|^{2},
\end{align}
where $\{|j\rangle\}_j ,\{p_j\}_j$ are the eigenstates and eigenvalues of $\rho$, respectively (the sum excludes any $p_j+p_k=0$).
For pure states, the QFI expression reduces to
$I_{Q}=4\left(\langle\partial_{\omega}\psi|\partial_{\omega}\psi\rangle-|\langle\psi|\partial_{\omega}\psi\rangle|^{2}\right).$ In the special case of $|\psi\left(\omega\right)\rangle=\exp\left(-i \mathcal{H} (\omega )T\right)|\psi\rangle$, which corresponds to a qubit in a pure state evolving under some Hamiltonian $\mathcal{H}(\omega)$ parametrized by $\omega$, the QFI is further simplified to $4T^{2}\text{var}_{|\psi\rangle}\left(\partial_\omega \mathcal{H}\right).$
In our case, $\mathcal{H}$ is the free evolution Hamiltonian, i.e. $\mathcal{H}=\omega \sigma_{z}$. Hence, the optimal QFI is
\begin{align}
\label{eq:qfi}
    I_{Q} = 4T^{2} \, \underset{|\psi\rangle}{\text{max}} \; \text{var}_{|\psi\rangle}\left(\sigma_{z}\right)=4T^{2}.
\end{align}
It can be shown that this is the optimal QFI when optimized over all possible control strategies and initial states \cite{pang_optimal_2017,zhou_asymptotic_2021}, hence the fundamental precision limit for this problem is $\Delta\omega\geq\frac{1}{2T}.$
For $N$ qubits, assuming only separable states and control, the limit is given by
$I_Q=4NT^2,$ hence
$\Delta\omega\geq\frac{1}{2\sqrt{N}T}.$ We refer to this bound as the QFI limit throughout the paper.

\subsection{Fisher Information Bounds for Weak Measurements Protocols}

Here, we 
derive the Cramér-Rao bounds for estimating the frequency 
$\omega$ using our weak measurement protocols. 
Since we consider a sequence of weak measurements on a qubit, the probability of obtaining an outcome $\vec{x}$ can be written as $p(\vec{x}|\omega) = \prod_{k=1}^{T/\tau} p(x_k|\omega)$, where $p(x_k|\omega)$ is given in Eq. (\ref{supp_eqn:prob_weak_meas}).
Note that the outcomes are correlated (since the outcome of the $k^{\text{th}}$ measurement changes the state of the $k+1$ measurement), which makes the analytical calculation of the CFI very challenging. We can however obtain expressions of the CFI in some regimes of the parameter space using relevant analytical bounds.

We study the CFI of two weak measurement protocols, which we refer to as the \text{weak-only} and the \textit{weak-with-strong} protocols. The former employs weak measurements with strength $g$ during the interrogation, where the interrogation time is $T$ and the measurement period is $\tau$. The latter also performs these weak measurements with the same measurement parameters, however swaps the last measurement (i.e. the $T/\tau^\text{th}$ measurement at time $t = T$) for a projective (strong) one, in which the sensing qubit is projected to the eigenstates of the Pauli $x$ operator $\sigma_x$.

We also define two regimes to describe the behavior of the CFI, characterized by the parameter $\eta = g^2T/\tau$. They are referred to as the \textit{weak back action} regime and the \textit{strong back action} regime, where $\eta \ll 1$ and $\eta \gg 1$, respectively. Since we have found that the phase contrast in the average dynamics decays with a rate of $\gamma=-\frac{1}{2\tau}\log\left(\cos\left(2g\right)\right) \approx g^2/\tau$ for $g\ll1$ (see Eq. (\ref{eq:avg_dynamics_eigenvals})), 
$\eta = T/\gamma^{-1}$ quantifies the ratio between the total interrogation time $T$ and the phase coherence decay time $\gamma^{-1}$.



\subsubsection{Upper bounds of the CFI}
A general upper bound to the CFI with sequential measurements of a probe was given in Ref. \cite{gammelmark_fisher_2014}.
This is done by considering the map:
\begin{align}
\Lambda\left(\omega_{1},\omega_{2}\right)\left(\rho\right):=K_{+}^{\left(\omega_{1}\right)}\rho K_{+}^{\left(\omega_{2}\right)\dagger}+K_{-}^{\left(\omega_{1}\right)}\rho K_{-}^{\left(\omega_{2}\right)\dagger},    
\end{align}
where $K_{\pm}^{\left(\omega_{i}\right)}=K_{\pm}V_{\omega_{i}}.$
Note that this is not a quantum channel, as it is not trace preserving (due to the different frequencies in $K_{\pm}^{\left(\omega_{i}\right)}$).
The bound for a time $T$, i.e. after $T/\tau$ applications of $\Lambda,$ is given by 
\begin{align}
I\leq4 \, \partial_{\omega_{1}}\partial_{\omega_{2}}\log\left(\text{tr}\left(\rho\left(\omega_{1},\omega_{2}\right)\right)\right) \mid_{\omega_{1}=\omega_{2}=\omega},
\label{eq:molmer_general_1}
\end{align}
where $\rho\left(\omega_{1},\omega_{2}\right)=\Lambda\left(\omega_{1},\omega_{2}\right)^{T/\tau}\left(\rho\right).$
It can be shown that this bound is simplified to
\begin{align}
I\leq4T\partial_{\omega_{1}}\partial_{\omega_{2}}\eta\mid_{\omega_{1}=\omega_{2}=\omega}
\label{eq:molmer_general_2}
\end{align}
where $\eta$ is the largest eigenvalue of $\Lambda\left(\omega_{1},\omega_{2}\right).$

\begin{figure}
    \centering
    \includegraphics[width=0.5\linewidth]{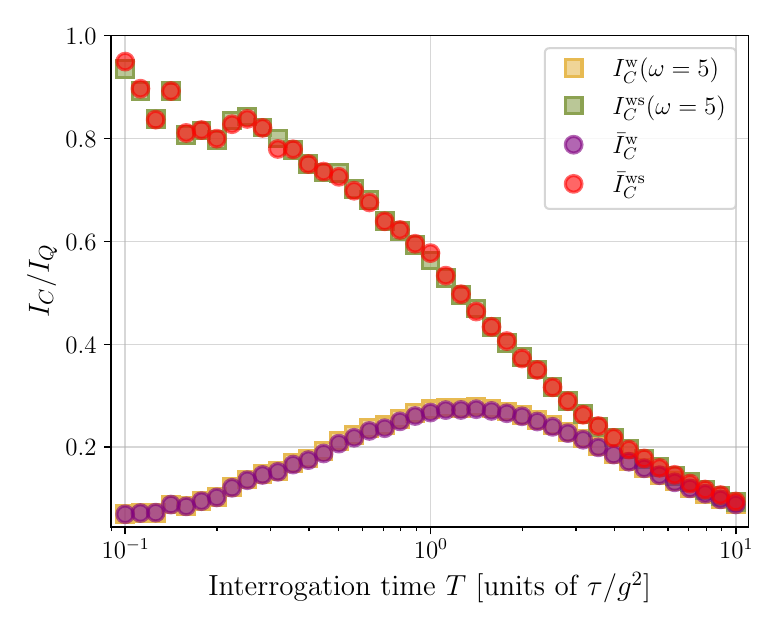}
    \caption{Comparison of the CFI $I_C$ (Eq. (\ref{eq:fi_defn})) and the average CFI $\bar I_C$ (Eq. (\ref{eq:avg_cfi_defn})) for the weak measurement protocols.
    We normalize both by the QFI, $I_Q=4NT^2$. We fix the measurement strength as $g=0.1$, measurement period as $\tau = 0.1$ s, and vary the interrogation time $T \in [1, 100]$ s. The CFI is calculated for $\omega = 5$ rad/s. We observe that in the limit of $\omega T \gg 1$, given that $\omega \tau < \pi/2$, $I_C \approx \bar I_C$ for both protocols.}
    \label{fig:cfi_avg_vs_cfi}
\end{figure}

\subsubsection{Expressions of the CFI}

We first study the behavior of the CFI in the weak back action regime, i.e.  
$\eta \ll 1$. 
In this regime, the back action due to weak measurements is negligible.
Therefore, measurement results for different times will approximately be independent of each other, and the total CFI can be written as a sum over the CFI obtained only from the $k^\text{th}$ measurement, in the form of $I_C = \sum_k I_C[k]$. Furthermore, in this limit, $p(x_k|\omega)$ for a weak measurement can simply be written as $ p(x_k| \omega) = \left[ 1 + (-1)^{x_k} \sin{(2g)} \cos{(2k \omega \tau)} \right]/2$, $x_k = 0, 1$. Therefore, the CFI for the $k^\text{th}$ weak measurement is calculated as
\begin{align}
    I_C[k] =  \frac{4\sin{(2g)}^2 k^2 \tau^2 \sin{(2 \omega k \tau)}^2}{1 - \sin{(2g)}^2 \cos{(2\omega k \tau)}^2} \approx 4\sin{(2g)}^2 k^2 \tau^2 \sin{(2 \omega k \tau)}^2    
\end{align}
Thus, the total CFI of the weak-only protocol for a single qubit is given by
\begin{align}
I_C^{\text{w}}=\sum_k I_C[k] \approx 2 \sin{(2g)}^2 T^3/3\tau \approx 8 g^2 T^3/3\tau,
\end{align}
for $T/\tau \gg 1$. For $N$ qubits, the total CFI is given by
\begin{align}
\label{eq:fisher_weak_only}
    I_C^{\text{w}} \approx \frac{8 g^2 N T^3}{3 \tau}
\end{align}

Now, let us find the CFI of the weak-with-strong protocol. In the regime where $\eta \ll 1$, the probability for the measurement outcomes of the projective measurement at the end of the interrogation is approximately given by 
\begin{align}
    p_s(x_{T/\tau}|\omega) &= \frac{1}{2} \left[1 + (-1)^{x_{T/\tau}} \left( \cos{(2\omega T)} + 2g \sin{(2 \omega T)} \sum\nolimits_{k = 1}^{T/\tau-1} (-1)^{x_k} \sin{( 2\omega k \tau)}  \right) \right] 
\end{align}
where $\vec{x} = (x_1, x_2, \dots, x_{T/\tau-1})$ are the measurement results of the weak measurements, and there is some back action on the qubit, on the first order of $g$. Then, the CFI of this protocol for a single qubit can be written as

\begin{align}
\label{eq:fi_strong}
   I_C^{\text{ws}} &= \frac{4T^2}{2^{T/\tau}} \sin{(2\omega T)} \sum_{\vec{x}} \frac{\left[1 - 2g \left(\cot{(2 \omega T)} \sum_{k = 1}^{T/\tau-1} (-1)^{x_k} \sin{( 2\omega k \tau)} - \frac{\tau}{T}  \sum_{k = 1}^{T/\tau-1} (-1)^{x_k} k \cos{( 2\omega k \tau)} \right) \right]^2}{1 + (-1)^{x_{T/\tau}}\sin{(2\omega T)} \left( \cot{(2\omega T)} + 2g \sum_{k=1}^{T/\tau-1} (-1)^{x_k} \sin{( 2\omega k \tau)} \right)} \nonumber \\
   & \approx  \frac{4T^2}{2^{T/\tau} } \sum_{\vec{x}} \left[ 1 - 2g \sum\nolimits_{k=1}^{T/\tau-1} (-1)^{x_k} \left( (\tan{(2\omega T)}-2\cot{(2\omega T)})  \sin{( 2\omega k \tau)} + \frac{2 \tau}{T} k \cos{( 2\omega k \tau)} \right)  \right]  \nonumber \\
     & \approx 4 T^2 + O(g^2)
\end{align}
We note that this protocol saturates the ultimate sensitivity limit with respect to the total interrogation time $T$ for a small measurement strength $g$, given by the QFI in Eq. (\ref{eq:qfi}). If we have $N$ qubits, we will similarly have that $I_C^{\text{ws}} \approx 4NT^2 + O(g^2)$.

In the strong back action regime, $\eta \gg 1$, the CFI of both protocols scales as $T$ due to phase coherence loss. 
The CFI in this regime can be evaluated using the upper bound of Eqs. (\ref{eq:molmer_general_1})-(\ref{eq:molmer_general_2}).
In our case $\partial_{\omega_{1}}\partial_{\omega_{2}}\eta\mid_{\omega_{1}=\omega_{2}=\omega}= \tau\cot\left(g\right)^{2}.$  This upper bound thus reads
\begin{align}
I \leq 4T \tau \cot\left(g\right)^{2}.    
\end{align}
Note that in general, $\tau\cot\left(g\right)^{2}\geq\gamma^{-1}=\left(-\frac{1}{2\tau}\log\left(\cos\left(2g\right)\right)\right)^{-1},$
but in the limit 
of $g \ll 1$ we have that $\gamma^{-1}\approx\tau\cot\left(g\right)^{2} \approx\frac{\tau}{g^{2}}$. Hence, in this limit,
\begin{align}
I\leq4T\gamma^{-1}\approx 4\frac{\tau}{g^{2}}T.
\label{eq:molmer_bound_strong_back action}
\end{align}
While this bound holds for every $T,$ it is tight only in the strong back action regime, i.e. $\eta \gg1.$
The intuition for this strong back action limit of the CFI is the following: For times shorter than $1/\gamma$, the frequency is sensed with Heisenberg scaling (HS), such that the CFI reaches $4/\gamma^2$. The measurements after $t=1/\gamma$ are uncorrelated with the measurements during time $t< 1/\gamma$. Thus, the CFI will be equivalent to the case where the measurement during $t<1/\gamma$ is repeated $T\gamma$ times, which is calculated as $4T/\gamma$. Note that the analysis for this upper bound of the CFI is not modified when we perform a strong measurement at the end of the interrogation, as we work in the regime that the strong measurement provides negligible information, since $\eta \gg 1$.

\begin{figure}
    \centering
\includegraphics[width=0.95\linewidth]{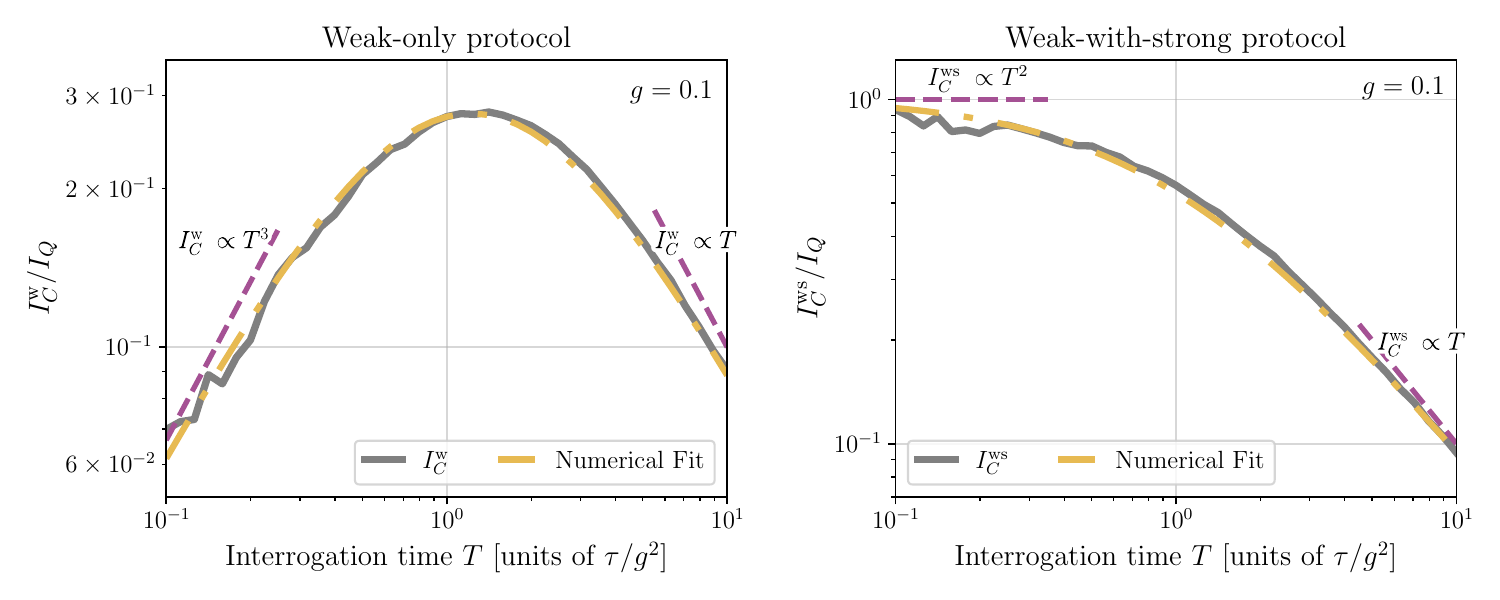}
    \caption{CFI, normalized by the QFI ($I_Q=4NT^2$), for the two weak measurement protocols. We set $g = 0.1$, $\tau = 0.1$ s, and $1< T < 100$ s, so that $g \ll 1$ and $T/\tau \gg 1$. Left: CFI for the weak-only protocol. We observe that the CFI scales as $T^3$ in the weak back action regime, where $\eta = g^2T/\tau \ll 1$. It scales as $T$ in the strong back action regime, where $\eta \gg 1$. The numerical fit plotted with the yellow dash-dotted line is given in Eq. (\ref{eq:fi_with_decay_weak_only}).
    The peak of $I_C^{\text{w}}/I_Q$ corresponds to $g^2T/\tau = \sqrt{3/2}.$ The CFI is maximized with respect to the measurement strength, $g$, at this point, and it also signifies the transition from the weak back action to the strong back action regime.
    Right: CFI for the weak-with-strong protocol. We observe that the CFI scales as $T^2$ in the regime where $\eta \ll 1$, and saturates the QFI $I_Q$.
    For $\eta \gg 1$, similar to the weak-only protocol, the CFI scales as $T$. The numerical fit plotted with the yellow dash-dotted line is given in Eq. (\ref{eq:fi_with_decay_strong}).}
    \label{fig:fi_scalings}
\end{figure}

We notice that the analytical expressions for the CFI for both protocols when $\eta \ll 1$ (Eqs. (\ref{eq:fisher_weak_only}) and (\ref{eq:fi_strong})) or $\eta \gg 1$ (Eq. (\ref{eq:molmer_bound_strong_back action})) are independent of $\omega$, in the limit of $T/\tau \gg 1$. Furthermore, we numerically observed that the CFI for both protocols becomes independent of $\omega$ when $\omega T \gg 1$, given that $\omega \tau <\pi/2$. This is demonstrated in Fig. \ref{fig:cfi_avg_vs_cfi}, where we plot $I_C^\text{w}$ and $I_C^\text{ws}$ for $\omega = 5$ rad/s, as well as $\bar I_C^\text{w}$ and $\bar I_C^\text{ws}$, as a function of the interrogation time $T$. We fix $\tau = 0.1$ s, $g = 0.1$, and vary $T \in [1, 100]$ s. For these parameters, we observe that $I_C$ converges $\bar I_C$ with increasing $T$. Since we focus on the extended dynamic range regime where $\omega T \gg 1$, $\omega \tau < \pi/2$, we therefore use the CFI $I_C$ and the average CFI $\bar I_C$ interchangeably for the rest of the Sections.

Numerically, we compute the CFI for a general protocol using a Monte Carlo sampling, i.e. by sampling $K$ trajectories of measurement outcomes for $N$ qubits. For large enough $K$, the CFI is approximately given by 
\begin{align}
I_C \approx \frac{1}{K} \sum_{\vec{x}} \left( \frac{\partial \ln{p(\vec{x}|\omega)}}{\partial \omega} \right)^2
\end{align}
where the sum is over the sampled trajectories $\vec{x}$, and $p(\vec{x}|\omega)$ is the probability of obtaining a trajectory $\vec{x}$ given $\omega$. We also numerically compute the derivative by calculating $p(\vec{x}|\omega+d\omega)$, $d\omega \ll 1$, for all $\vec{x}$.

\subsubsection{Numerical Fits for the Fisher Information}

From the analysis above, we obtained analytical expressions of the CFI for the weak-only protocol, in the limit of $g \ll1$, in two asymptotic regimes:
\begin{align}
I_C^{\text{w}} = \begin{cases}
\frac{8}{3}Ng^{2}\frac{T^{3}}{\tau} & \eta\ll1\\[2mm]
\frac{4NT\tau}{g^{2}} & \eta\gg1
\end{cases}.
\end{align}
These regimes correspond to the weak and strong back action limits, respectively.
The weak back action expression was derived in 
Eq. (\ref{eq:fisher_weak_only}), and the strong back action expression is based on the upper bound of Eq. (\ref{eq:molmer_bound_strong_back action}) which appears to be tight from our numerical simulations. We expect the full expression for the CFI to interpolate between these two limits. 
With this constraint, we obtain a good fit to the CFI numerically for $g \ll 1$:
\begin{align}
\label{eq:fi_with_decay_weak_only}
    I_C^{\text{w}} \approx \frac{8 g^2 N T^3}{3\tau} \cdot \frac{1}{1 + 0.77 \, g^2 \frac{T}{\tau} + \frac{2g^4}{3} \frac{T^2}{\tau^2}}.
\end{align}
The CFI is maximized with respect to the measurement strength, $g$, when $g^2T/\tau = \sqrt{3/2}$. For this value of the measurement strength, the CFI reaches its maximum value, $I_C^{\text{w}} \approx 1.11 NT^2$.

\begin{figure}
    \centering\includegraphics[width=0.5\linewidth]{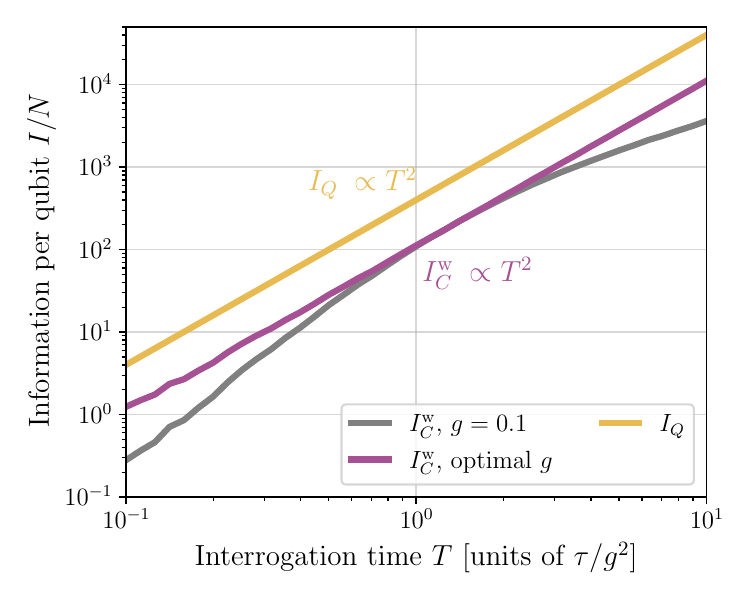}
    \caption{CFI for the weak-only protocol, for optimal measurement strength $g$. We choose the measurement period as $\tau = 0.1$, and vary the total interrogation time in the range of $1 < T< 100$ s.
    We plot the CFI for a fixed measurement strength, $g= 0.1$, with a solid, gray line. This is the measurement strength used to denote the units of the interrogation time $T$ in the x-axis. Then, we apply the measurement strength that maximizes the CFI, i.e. $g = \sqrt{\tau/T} \, (3/2)^{1/4}$, and plot the CFI with a solid, purple line. For this measurement strength, the CFI shows HS, scaling as $\approx 1.11 NT^2$. Finally, we plot the QFI limit for reference with a solid, yellow line. Note that even though we obtain HS, we do not saturate the QFI for this protocol.}
\label{fig:fi_weak_optimal_g}
\end{figure}

\begin{figure}
    \centering
\includegraphics[width=0.5\linewidth]{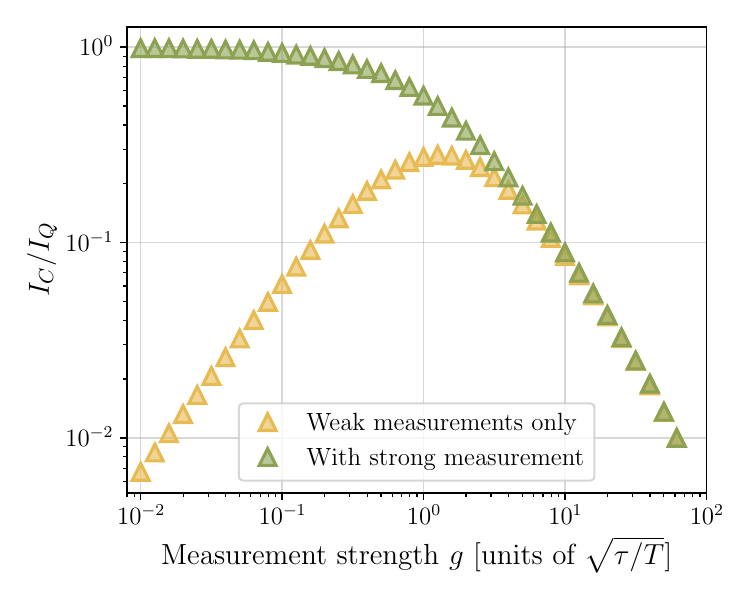}
    \caption{Classical Fisher information (CFI) $I_C$, normalized by the QFI, $I_Q=4NT^2$, for the weak-only protocol (plotted in yellow), and weak-with-strong protocol (plotted in green). We fix the measurement period as $\tau = 0.1$ s, the total duration of the experiment as $T = 10$ s, the  frequency as $\omega = 5$ rad/s, and vary the measurement strength $g \in [0.01, \pi/4]$. }
    \label{fig:fi_wrt_g}
\end{figure}

Similarly, for the weak-with-strong protocol, the analytical expressions for the CFI in the two asymptotic regimes are computed as (for $g \ll 1$)
\begin{align}
I_C^{\text{ws}} = \begin{cases}
4NT^2 & \eta\ll1\\[2mm]
\frac{4NT\tau}{g^{2}} & \eta\gg1
\end{cases}.
\end{align}
Here, the CFI decreases monotonously as $g$ increases. Furthermore, the CFI scales as $T^2$ for $\eta \ll 1$, and $T$ for $\eta \gg 1$, respectively.
Numerically, we observed that a good fit is for the CFI for $g \ll 1$ is given by
\begin{align}
\label{eq:fi_with_decay_strong}
    I_C^{\text{ws}} \approx \frac{4NT^2}{1 - 0.13 \,g \, \sqrt{\frac{T}{\tau}} + g^2\frac{T}{\tau} } \text{ for } g \ll 1.
\end{align} 
We perform numerical simulations to compute the CFI for both protocols in Figs. \ref{fig:fi_scalings}, \ref{fig:fi_weak_optimal_g}, and \ref{fig:fi_wrt_g}. First, in Fig. \ref{fig:fi_scalings}, we set the measurement strength as $g = 0.1$, and sample the unknown frequency $\omega$ from the uniform prior distribution in $[0, \delta\omega]$, where we set $\delta\omega = 5\pi$ rad/s. $\delta\omega$ determines the necessary measurement period for the weak measurements, i.e. $\tau = \pi/2\delta\omega = 0.1$ s.
We vary the total interrogation time in the range of $1 < T < 100$ s, such that $g \ll 1$ and $T/\tau \gg 1$. 
The numerical results for the CFI are plotted in plain gray lines, whereas the numerical fits given in Eqs. (\ref{eq:fi_with_decay_weak_only}) and (\ref{eq:fi_with_decay_strong}) are plotted in yellow dash-dotted lines. 
Then, we re-plot the CFI for the weak-only protocol in Fig. \ref{fig:fi_weak_optimal_g}, however, we vary the measurement strength as a function of the total interrogation time $T$. More specifically, we select the measurement strength that maximizes the CFI, i.e. $g = \sqrt{\tau/T} \, (3/2)^{1/4}$ for all $T$ (see Eq. (\ref{eq:fi_with_decay_weak_only})). The rest of the measurement parameters $T, \tau, \delta\omega$ are unchanged, and can be found in the paragraph above. We observe that since the CFI is optimized with respect to $g$, it is given by $I_C^{\text{w}} \approx 1.11 NT^2$ for all $T$. Therefore, it shows HS with respect to the total interrogation time, but does not saturate the QFI, given by $I_Q = 4NT^2$. 
Finally, we plot the CFI of both protocols as a function of the measurement strength $g$ in Fig. \ref{fig:fi_wrt_g}. We set $T = 10$ s and $\tau = 0.1$ s, such that we operate in the extended dynamic range regime. As predicted analytically, in the limit of $\eta \ll 1$, the CFI of the weak-with-strong protocol is independent of $g$, whereas that of the weak-only protocol scales as $g^2$. In the limit of $\eta \gg 1$, the CFI of both protocols coincide and decrease as $g^{-2}$ for $g \ll 1$.

\section{ Threshold Derivation: Analytical Model for the Maximum Likelihood Estimator}

In order to derive the variance of the maximum likelihood estimator (MLE), we work in the regime where the number of qubits is large ($N \gg 1$), the measurement strength is small ($g^2 
< \tau/T$), and the number of measurements is large $T/\tau \gg 1$. In this regime, the back action due to weak measurements is negligible; hence, subsequent measurements 
can be treated as approximately independent events. Let us denote the measurement results in the computational basis $\{ \ket{0}, \ket{1}\}$ for a time $t = n\tau$ as $y_n=0,1$, and $x_+(n)$ stands for the number of the $y_n=0$ outcomes normalized by the number of qubits, $N$. 
Denoting the probability of $y_n=0$ as $p(y_n=0|\omega)$, then in the limit of $N \gg 1$, $N p(y_n=0|\omega),N p(y_n=1|\omega) \gg 1,$
we can 
use the central limit theorem and approximate the probability density function of $x_+(n)$ as
\begin{align}
\label{eq:binomial_gaussian}
    P\left(x_+(n) |\omega\right) \sim \frac{\sqrt{N}}{\sqrt{2\pi p(y_n=0|\omega) (1-p(y_n=0|\omega))}} \text{exp}\left(- \frac{N( x_+(n) - p(y_n=0|\omega))^2}{2 p(y_n=0|\omega) (1-p(y_n=0|\omega))}\right)
\end{align}
Here, we used the normal approximation to the binomial distribution, such that $x_+(n)$ is a Gaussian random variable with $E[x_+(n)] = p(y_n=0|\omega)$, and $\text{var}(x_+(n)) = p(y_n=0|\omega) (1-p(y_n=0|\omega))/N$. Note that in this weak back action regime $p(y_n|\omega)$ is given by $p(y_n|\omega) = \left[ 1 + (-1)^{y_n}\sin{(2g)} \cos{(2\omega n \tau)} \right]/2$. Then, $E[x_+(n)] = \left[ 1 + \sin{(2g)}\cos{(2\omega n\tau)}\right]/2 \approx  1/2 + g \cos{(2\omega n\tau)}$, and $\text{var}(x_+(n)) \approx 1/4N$. 
In this limit the problem is thus equivalent to the problem of classical signal estimation \cite{rife_single_1974}:
we sample a classical signal $s\left(t_{n}\right)=\frac{1}{2}\left(1+\sin\left(2g\right)\cos\left(2\omega t_{n}\right)\right)+\nu_{n}$ where $\left\{ \nu_{n}\right\} _{n=1}^{T/\tau}$ are i.i.d Gaussian random variables $\nu_{n}\sim N\left(0,\frac{1}{4N}\right)$ and $t_{n}=n\tau$ .

The MLE is therefore given by 
$\text{argmax}_\omega\left[\log{\prod_{n=1}^{T/\tau}P(x_{+}(n)|\omega)}\right].$
Using the probability distribution for $x_+(n)$ given in Eq. (\ref{eq:binomial_gaussian}), we can write this expression as
\begin{align}
    \omega_{\text{MLE}}  \coloneqq \underset{\omega}{\text{argmin}}\sum_{n=1}^{T/\tau} \left(x_+\left( n \right) - p(y_n=0|\omega) \right)^2 = \underset{\omega}{\text{argmin}}\sum_{n=1}^{T/\tau} \left(x_+\left( n \right) - \frac{1}{2} \left( 1 + \sin{(2g)} \cos{(2\omega n \tau)} \right) \right)^2.
    \label{eq:mle_1}
\end{align}
For brevity, hereafter we will replace $x_+\left( n \right)$ with the more compact notation $x_n.$
We expand the expression of Eq. (\ref{eq:mle_1}) and discard the $x_n^2$ terms. In the limit that $g \ll 1$, the minimization above can approximated as 
\begin{align}
\label{eq:mle_simple_defn}
    \omega_{\text{MLE}} \approx \underset{\omega}{\text{argmax}} \sum_{n=1}^{T/\tau}  x_n \cos{(2\omega n \tau)} 
\end{align}
The expression inside the maximization is the definition of the real part of a discrete time Fourier transform (DTFT), which is a continuous function of frequency. Since we work in the limit that $T \gg \tau$, the DTFT is closely linked to the discrete Fourier transform (DFT), i.e. we do not lose significant information by working with the discrete frequency bins instead of maximizing a continuous function. We can define the DFT operation as (matching the convention of the previous equations)
\begin{align}
\label{eq:A_k_defn}
    A_k = \frac{\tau}{T}\sum_{n=1}^{T/\tau} x_{n} \, \text{exp}\left(i \frac{2 \pi k n \tau}{T} \right), \; k = 1, 2, \dots, T/\tau, \quad x_{n} = \sum_{k = 1}^{T/\tau} A_k \, \text{exp}\left(-i \frac{2 \pi k n \tau}{T}  \right)
\end{align}
Then, we see that $\omega_{\text{MLE}} \approx \pi/T \, \text{argmax}_k \, \text{Re}(A_{k}) = \pi/T \, \text{argmax}_k \, B_k$, where we denoted the real part of $A_k$ as $B_k$. 
Note that while the DFT frequencies are $\left\{ \omega_{k}|\omega_{k}=\frac{\pi}{T}k\right\} _{k=1}^{T/\tau},$
in our case, 
because of the symmetry of the cosine function in Eq. (\ref{eq:mle_simple_defn}), we cannot distinguish between 
$\omega_{k}=\frac{\pi}{T}k$ and $\omega_{M-k}=\frac{\pi}{T}(m-k).$
The likelihood function thus has this symmetry which means that there will be two identical global maximas: $\omega_{k}$ and $\omega_{m-k}.$ 
This is however not an issue for us since the prior frequency distribution corresponds to $\left[0,\pi\frac{m}{2}\right]$ (where $m$ is the total number of measurements $m = \lfloor T/\tau\rfloor$),
therefore the maximization with respect to $\omega$ is only over the range $\left[0,\pi\frac{m}{2}\right],$
which does not suffer from this symmetry.
For simplicity, in what follows we will consider maximization over all of $\left\{ \omega_{k}|\omega_{k}=\frac{\pi}{T}k\right\} _{k=1}^{T/\tau},$
where we understand that the maximum corresponds to a pair of frequencies from which the estimator is the frequency in the range of $\left[0,\pi\frac{m}{2}\right].$

We observe from numerics that the variance of the MLE has a strong dependence on the signal-to-noise-ratio (SNR).
Here, as the amplitude of the signal is given by $\sin{(2g)} \approx O(g)$, $\text{SNR} \approx O(g^2 N T/\tau).$
For large enough SNR 
the MSE converges to the CRB, i.e. $(\Delta \omega_{\text{MLE}})^2 \approx I^{-1}_C.$
Conversely, for small SNR, the likelihood is very close to a uniform distribution over 
$[0, \pi/2\tau]$. Therefore, 
the variance of the MLE will approximately be $\pi^2/48\tau^2$. This idea can be summarized in the form of
\begin{align}
\label{eq:var_mle}
    (\Delta \omega_{\text{MLE}})^2 \approx \text{Prob}(\text{no outlier}) \, \frac{1}{I_C} + \text{Prob}(\text{outlier}) \frac{\pi^2}{48\tau^2}
\end{align}
where the probability of having no outliers corresponds to the large SNR case mentioned above. Let us denote it as $1-q$. We can write down this probability as
\begin{align}
\label{eq:outlier_prob_defn}
    1-q = \text{Prob}(B_k < B_{k^*} \; \forall \; k \neq k^*)
\end{align}
where $k^*$ is the frequency bin that is associated with the true value of the frequency, and $B_k, k \neq k^*$ are the rest of the frequency bins. Therefore, an outlier occurs if the noise in one of these frequency bins exceeds the signal in the bin $k^*$. Without loss of generality, we can assume that $\omega^* = \pi/4\tau$ is the frequency that we are trying to estimate, which is approximately the mean value of all possible true frequencies. Then, the respective frequency bin is $k^* = \lfloor T/4\tau \rceil$, or $k^* = \lfloor 3T/4\tau \rceil$, where we define $\lfloor x \rceil$ as the nearest integer to the real number $x$. For simplicity, let us assume that $\text{mod}(T, 4\tau) = 0$, so that $T/4\tau$ is an integer. The fact that there can be two correct frequency bins is due to the symmetry of the cosine in the definition of the MLE, also mentioned above. Then, $x_n$ is a Gaussian random variable with $E[x_n]  \approx  1/2 + g \cos{(\pi n/2)}$, and $\text{var}(x_n)\approx 1/4N$. From the definition of $B_k$,
\begin{subequations}
\label{eq:B_k_defn}
\begin{align}
    B_k &= \frac{\tau}{T}  \sum_{n=1}^{T/\tau} x_n \cos\left(2 \pi k n \tau/ T \right)  \\
    &=\frac{\tau}{T} \left( g \sum_{n=1}^{T/\tau} \cos{(\pi n/2)} \cos\left(2 \pi k n \tau/ T \right) +  \sum_{n=1}^{T/\tau} \nu_n  \cos\left(2 \pi k n \tau/ T \right) \right)
\end{align}
\end{subequations}
where $\{ \nu_n \}$, $n = 1, 2, \dots, T/\tau$ are independent zero-mean Gaussian random variables with a variance of $1/4N$. Since we assumed that $T/4\tau$ is an integer, the first term in this expression simplifies to
\begin{align}
    \sum_{n=1}^{T/\tau} \cos{(\pi n/2)} \cos\left(2 \pi k n \tau/ T \right) = \begin{cases}
        0, &\text{ if } \; k \in \{1, 2, \dots T/\tau \} \backslash \{ \frac{T}{4\tau}, \frac{3T}{4\tau} \}  \\
       \frac{T}{2\tau}, &\text{ if } \; k \in \{ \frac{T}{4\tau}, \frac{3T}{4\tau} \}
    \end{cases}
\end{align}
We then see that
\begin{align}
    B_k \sim 
    \begin{cases}
        \mathcal{N}(0, \tau/8NT) & \text{if} \; k \neq \frac{T}{4\tau}, \frac{3T}{4\tau}\\
        \mathcal{N}(g/2, \tau/8NT) & \text{otherwise}.
    \end{cases}
\end{align} 

In order to calculate the outlier probability of Eq. (\ref{eq:outlier_prob_defn}), we also need to inquire whether there is a dependency between the variables $B_k$. Using the definition of $B_k$ in Eq. (\ref{eq:B_k_defn}), we can write down the normalized cross-correlation between $B_{k_1}$ and $B_{k_2}$, $k_1, k_2 = 1, 2, \dots, T/\tau$,
\begin{align}
    \rho_{B_{k_1} B_{k_2}} = \frac{E[(B_{k_1}-E[B_{k_1}]) (B_{k_2}-E[B_{k_2}])]}{\sigma_{B_{k_1}} \sigma_{B_{k_2}}} = \begin{cases}
        0 & \text{ if } k_1 \neq k_2, \; k_1 \neq T/\tau - k_2 \\
        1 & \text{ otherwise}.
    \end{cases}
\end{align}
where $\sigma_{B_{k_1}} = \sqrt{\tau/8NT}$ is the standard deviation of $B_{k_1}$. Therefore, the total number of statistically uncorrelated frequency bins is $T/2\tau$. By using these independent frequency bins, we can redefine the probability of no outliers occurring in the MLE (in Eq. (\ref{eq:outlier_prob_defn})) as
\begin{align}
\label{eq:outlier_prob_defn_2}
    1 - q = \text{Prob}(B_k < B_{T/4\tau} \; \forall \, k \in \{1, 2, \dots, T/2\tau \} \backslash \{T/4\tau \} ) 
\end{align}
Since $B_k$ for $k\neq T/4\tau$ are identically distributed, we can write this expression as
\begin{subequations}
\begin{align}
    1 - q &= \int dx \, \text{Prob}(B_1 < x )^{T/2\tau - 1} \, \text{Prob}( B_{T/4\tau} = x ) \\
    &= \frac{1}{\sqrt{2^{\frac{T}{\tau}-1}\pi \sigma^2}} \int_{0}^\infty dx \left[\left( 1 + \text{erf}\left(\frac{x}{\sqrt{2 }\sigma} \right) \right)^{\frac{T}{2\tau}-1}  e^{-\frac{(x - \mu)^2}{2\sigma^2}} +\left( 1 - \text{erf}\left(\frac{x}{\sqrt{2 }\sigma} \right) \right)^{\frac{T}{2\tau}-1}  e^{-\frac{(x + \mu)^2}{2\sigma^2}} \right]
\end{align}
\end{subequations}
where $\mu = g/2$, and $\sigma^2 = \tau/8NT$, and $\text{erf}(x)$ is the error function defined with $\text{erf}(x) = \frac{2}{\sqrt{\pi}}\int_0^z dt \, e^{-t^2}$. Let us assume that  $g^2NT/\tau \gg 1$, i.e. $\mu/\sigma \gg 1$. In this limit, we expect the outlier probability to be small, as the SNR is large. Furthermore, the largest contribution to the integral will be around $x \approx \mu$, due to the exponential factor in the first term. The second term will peak at $x \approx -\mu$, which is not within the integration limits. Therefore, we expect this term to decrease monotonously as the integration variable $x$, increases, and we can ignore this term as it will be very small compared to the first term. Therefore, we have
\begin{subequations}
\begin{align}
    1 - q &\approx \frac{1}{\sqrt{2 \pi \sigma^2}}\int_{0}^\infty dx \left(1 - \frac{1}{2}\text{erfc}\left(\frac{x}{\sqrt{2 }\sigma} \right) \right)^{\frac{T}{2\tau}-1}  e^{-\frac{(x - \mu)^2}{2\sigma^2}}  \\
    &\approx \frac{1}{\sqrt{2\pi \sigma^2}}\int_{0}^\infty dx  \left(1 - \frac{T}{4\tau}\text{erfc}\left(\frac{x}{\sqrt{2 }\sigma} \right) \right)  e^{-\frac{(x - \mu)^2}{2\sigma^2}} \label{eq:erfc_approx} \\
    &\approx 1 - \frac{1}{\sqrt{2\pi\sigma^2}}\frac{T}{4\tau} \int_0^\infty dx \, \text{erfc}\left(\frac{x}{\sqrt{2 }\sigma} \right) e^{-\frac{(x - \mu)^2}{2\sigma^2}}
    \label{eq:erfc_approx_2}
\end{align}
\end{subequations}
where $\text{erfc}(x)=1-\text{erf}(x)$ is the complementary error function. The approximation in Eq. (\ref{eq:erfc_approx}) holds when $\text{erfc}\left(x/\sqrt{2 }\sigma\right) T/\tau \ll 1$ for $x$ near $\mu$. Since we assumed that $\mu/\sigma \gg 1$, we can approximate the error function as
$\text{erfc}\left(x/\sqrt{2}\sigma \right) \approx  \,e^{-x^2/2\sigma^2}\sqrt{2}\sigma/\sqrt{\pi}x (1 + O(x^{-2}))$. Then, Eq. (\ref{eq:erfc_approx}) holds when
\begin{align}
    \text{erfc}\left(\frac{\mu}{\sqrt{2 }\sigma} \right) \frac{T}{\tau} \approx e^{-g^2NT/\tau} \sqrt{\frac{T}{g^2\pi N\tau}} \ll 1,
\end{align}
or, when $g^2NT/\tau \gg \log{(T/\tau)}$. Since $T/\tau \gg 1$, $T/\tau \gg \log{(T/\tau)}$, therefore this approximation holds when $g^2 \approx O(1/N)$ or larger. Furthermore, in Eq. (\ref{eq:erfc_approx_2}), we integrated over the constant term by extending the lower integration limit to $-\infty$, which is a good approximation when $\mu/\sigma \gg 1$. We can finally use the approximated complementary error function to compute $1-q$:
\begin{align}
    1-q \approx 1 - \frac{T}{4\pi\tau} e^{-\frac{\mu^2}{4\sigma^2}} \int_\epsilon^\infty dx \, \frac{1}{x} e^{-\frac{(x - \mu/2)^2}{\sigma^2}}
\end{align}
where $\epsilon$ is a small number close to zero to make sure that the integral converges. We employ Laplace's method to compute this expression for $\mu/\sigma \gg 1$. The integrand will peak near $x \approx \mu/2$, therefore we write
\begin{align}
    1-q \approx 1 - \frac{T}{4\pi\tau} e^{-\frac{\mu^2}{4\sigma^2}} \frac{2}{\mu} \int_0^\infty dx \, e^{-\frac{(x - \mu/2)^2}{\sigma^2}} \approx 1 - \frac{T}{2\sqrt{\pi}\tau} \frac{\sigma}{\mu} e^{-\frac{\mu^2}{4\sigma^2}}
\end{align}
Plugging in $\mu$ and $\sigma$, we obtain for the probability of having no outliers
\begin{align}
\label{eq:prob_no_outliers_final}
    1-q = 1 - \sqrt{\frac{T}{8\pi N \tau g^2}} e^{-g^2NT/2\tau}
\end{align}
Therefore, given this probability, and the CFI of the weak-only protocol in Eq. (\ref{eq:fisher_weak_only}), the variance of the MLE given in Eq. (\ref{eq:var_mle}) can be computed as
\begin{align}
\label{eq:var_mle_weak_only}
    (\Delta\omega_{\text{MLE}})^2\approx \left( 1 - \sqrt{\frac{T}{8\pi N \tau g^2}} e^{-g^2NT/2\tau} \right) \, \frac{3\tau}{8g^2NT^3} + \sqrt{\frac{T}{8\pi N \tau g^2}} e^{-g^2NT/2\tau} \frac{\pi^2}{48\tau^2}
\end{align}
We want to find the region of the parameter space where the information obtained from the MLE is very close to the CFI. We can express this with the parameter $\epsilon$, and write $I_\text{MLE} = (1-\epsilon) \, I_C^{\text{w}}$, where $\epsilon \ll 1$, and $I_\text{MLE} = 1/\text{var(MLE)}$. From Eq. (\ref{eq:var_mle_weak_only}), $1-\epsilon$ is given by
\begin{subequations}
\begin{align}
    1-\epsilon = \left[ 1 - q  + q \frac{\pi^2 g^2 N}{18} \left( \frac{T}{\tau}\right)^3\right]^{-1} &\approx 1 - q \frac{\pi^2 g^2 N}{18} \left( \frac{T}{\tau}\right)^3 \\
    &= 1 - \frac{\pi^{3/2}}{36} \sqrt{\frac{g^2NT}{2\tau}} \left( \frac{T}{\tau} \right)^3 e^{-g^2NT/2\tau}
\end{align}
\end{subequations}
where $1-q$ is the probability of having no outliers in Eq. (\ref{eq:prob_no_outliers_final}). The approximation holds when $g^2N (T/\tau)^3 \gg 1$. Since we already assumed that $g^2NT/\tau \gg 1$ while deriving the probability of having no outliers, and since $T/\tau \gg 1$, this is a valid approximation. Rearranging this equation, we obtain
\begin{align}
\label{eq:transcendental}
    e^x \approx \frac{\sqrt{x}}{\epsilon}  \frac{\pi^{3/2}}{36} \left(\frac{T}{\tau}\right)^3
\end{align}
where we defined $x = g^2NT/2\tau$. This is a transcendental equation, therefore we will find an approximate solution. Since $x \gg 1$, $x \gg \ln{x}$, and we have
\begin{align}
\label{eq:soln_1_transcendental}
    x \approx \ln{\left(\frac{\pi^{3/2}}{36\epsilon} \left(\frac{T}{\tau}\right)^3 \right)} + \frac{1}{2}\ln{(x)} \approx \ln{\left(\frac{\pi^{3/2}}{36\epsilon} \right)} + 3\ln{\left(\frac{T}{\tau}\right)}
\end{align}
To find the first order correction, let us write $x = x^* + \Delta x$, where $x^*$ is given in Eq. (\ref{eq:soln_1_transcendental}), and assume that $\Delta x \ll x^*$. Plugging this into Eq. (\ref{eq:transcendental}), and taking the logarithm of both sides, we find
\begin{align}
    \Delta x = \frac{1}{2} \ln{(x^* + \Delta x)} \approx \frac{1}{2} \ln{(x^*)} \; \text{ if } x^* \gg \Delta x
\end{align}
Therefore, the second order approximation to Eq. (\ref{eq:transcendental}) is given by (we also plug in the definition of $x$):
\begin{align}
\label{eq:threshold_cond_1}
    \frac{g^2NT}{2\tau} \approx \ln{\left(\frac{\pi^{3/2}}{36\epsilon} \right)} + 3\ln{\left(\frac{T}{\tau}\right)} + \frac{1}{2} \ln{\left[\ln{\left(\frac{\pi^{3/2}}{36\epsilon} \right)} + 3\ln{\left(\frac{T}{\tau}\right)} \right]}
\end{align}
Given the value for $\epsilon$, and some of the system parameters, this solution defines the point at which the ratio between the information obtained from the MLE and the maximum obtainable information is $1-\epsilon \approx 1$, as a function of a free system parameter. For example, we can fix the total interrogation time $T$, measurement period $\tau$, and the number of qubits $N$, and find the minimum measurement strength $g$ necessary to surpass this threshold (and obtain a larger information from the MLE). Or, we can fix the parameters $g$, $T$, and $\tau$ to find the necessary number of qubits to reach this this threshold. In order to operate in the favorable region where the CFI is maximized with respect to the measurement strength, $g$, we need to have $g^2T/\tau \approx \sqrt{3/2} \sim O(1)$ for the weak-only protocol (see Eq. (\ref{eq:fi_with_decay_weak_only})). Then, from Eq. (\ref{eq:threshold_cond_1}), we have that the number of qubits to reach this threshold scales as $N \sim O(3\ln{(T/\tau)})$.

Up to this point, we have only performed weak measurements on the qubits in order to estimate the frequency $\omega$. Now, let us perform a strong projective measurement at the end of the interrogation, in addition to the weak measurements (i.e. apply the weak-with-strong protocol). For this case, the total CFI will be the sum of the CFI due to the weak and the strong measurement, which were computed to be $8Ng^2T^3/3\tau$, and $4NT^2$, respectively. Note that the total achievable CFI for this case is larger than the QFI, $4NT^2$, as we have neglected the back action due to the weak measurements.

Furthermore, we can still define the variance of the MLE with Eq. (\ref{eq:var_mle}). Then, we need to find again the probability of having no outliers, $1-q$. For this purpose, let us redefine the MLE:
\begin{subequations}
\begin{align}
    \omega_{\text{MLE}} &\coloneqq  \underset{\omega}{\text{min}}\sum_{n=1}^{T/\tau-1} \left(x_n - \frac{1}{2} \left( 1 + \sin{(2g)} \cos{(2\omega n \tau)} \right) \right)^2 + \left(x_{T/\tau} - \frac{1}{2} \left( 1 + \cos{(2\omega T)} \right) \right)^2 \\
    &\approx  \underset{\omega}{\text{max}} \sum_{n=1}^{T/\tau-1} \left( 2g \, x_n \, \cos{(2\omega n \tau)} \right) + x_{T/\tau} \, \cos{(2\omega T)} - \cos{(\omega 
    T)}^4
\end{align}
\end{subequations}
where we ignored the term proportional to $g^2 T/\tau$ as $g \ll 1$, and the remaining terms are $O(g T/\tau)$ and $O(1)$, respectively. We can again define
\begin{align}
    B'_k = \frac{\tau}{T}\left(\sum_{n=1}^{T/\tau -1} x_{n} \, \text{cos}\left(2 \pi k n \frac{\tau}{T} \right) + \frac{1}{2g} x_{T/\tau} \cos{\left(2\pi k \frac{T}{\tau}\right)}\right)
\end{align}
where we have defined $B_k$ as the real part of $A_k$ in Eq. (\ref{eq:A_k_defn}), and $k = 1, 2, \dots, T/\tau$. Then, in the limit that $T/\tau \gg 1$, the optimization of the MLE reduces to the following: $\omega_{\text{MLE}} \approx \text{max}_k \; 2g B\, _k' - \cos{(\pi k T/\tau)}^4 \tau/T$. As before, let us assume that the true value of the frequency $\omega$ is $\omega^* = \pi/4\tau$, approximately the mean value of possible (detectable) frequencies. Furthermore, let us assume that $\text{mod}(T, 4\tau) = 0$, such that the true value of the frequency corresponds to an integer value of $k$. Then, the MLE problem is modified as
\begin{align}
    \omega_{\text{MLE}} = \underset{k}{\text{max}} \; 2g B\, _k' -  \tau/T =  \underset{k}{\text{max}} \; \sum_{n=1}^{T/\tau -1} x_{n} \, \text{cos}\left(2 \pi k n \frac{\tau}{T} \right)
\end{align}
as $\cos{(\pi k T/\tau)} = \cos{(2\pi k T/\tau)} = 1$ for $k \in \mathbb{N}$. Then, we realize that the MLE is almost exactly the same as the case where we only had weak measurements, with the difference being that the summation in that case also included the $T/\tau^\text{th}$ sample, measured with strength $g$ instead of 1. However, in the limit that $T/\tau \gg 1$, this difference is negligible. Therefore, we can employ the same expression for the probability of having no outliers as done for the case where we had weak measurements only (Eq. (\ref{eq:prob_no_outliers_final})).

\begin{figure}
    \centering
\includegraphics[width=\linewidth]{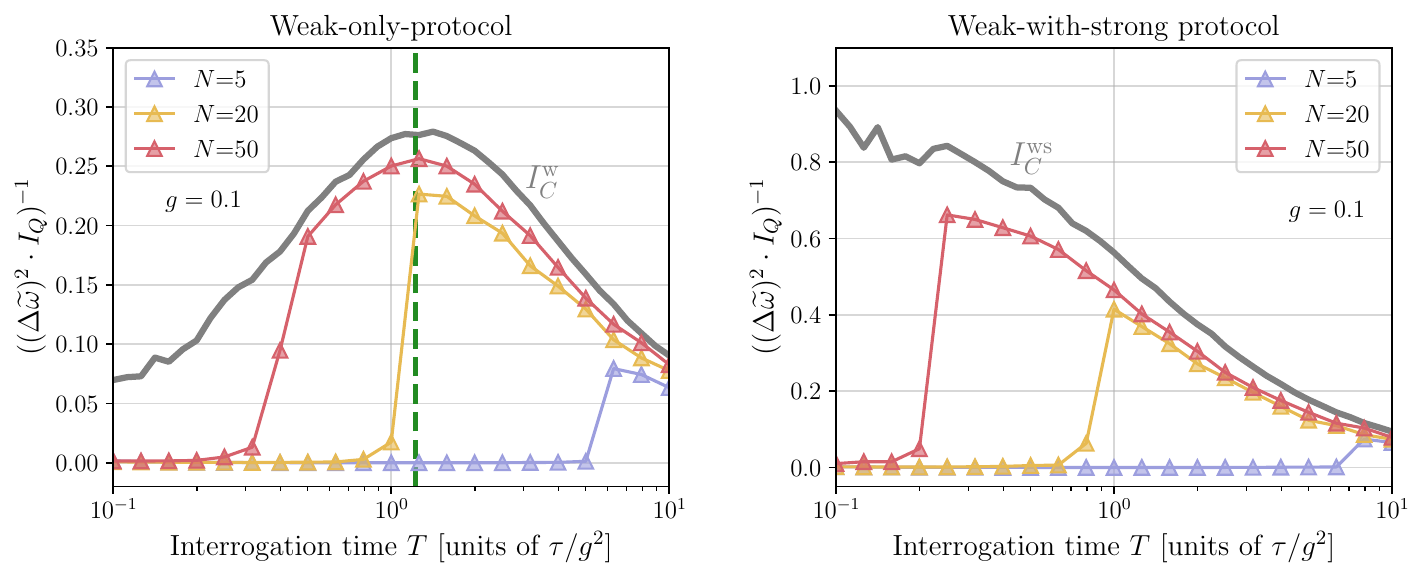}
    \caption{Performance of the weak-only protocol (left) and weak-with-strong protocol (right) with respect to the interrogation time $T$. We plot the inverse of the MSE scaled by the QFI $I_Q$, which quantifies the amount of information gained from the estimation. We fix the measurement strength as $g = 0.1$, and sample the unknown frequency from the uniform prior distribution in $[0, \delta\omega]$, where $\delta\omega = 5\pi$, such that the measurement period is chosen as $\tau = \pi/2\delta\omega=0.1$ s. Finally, we plot the CFI of both protocols, $ I_C^{\text{w}}$ and $ I_C^{\text{ws}}$, normalized by the QFI, as the benchmark. The CFIs are about $\omega = 5$ rad/s. The information gained from the estimation does not saturate the CFI for small SNR (i.e. small $T$) due to the threshold effect. As the SNR is proportinal to the number of qubits $N$, we can saturate to the CFI faster (for smaller $T$) with a larger number of qubits. In order to obtain HS, this saturation needs to occur in the weak back action regime: the transition between the two regimes is marked by a green dashed line in the plot on the left. }
    \label{fig:mse_wrt_T}
\end{figure}

Thus, the variance of the MLE can be written as
\begin{align}
\label{eq:var_mle_with_strong}
    \text{var}(\text{MLE}) \approx \left( 1 - \sqrt{\frac{T}{8\pi N \tau g^2}} e^{-g^2NT/2\tau} \right) \left[\frac{8Ng^2T^3}{3\tau} + 4NT^2\right]^{-1} + \sqrt{\frac{T}{8\pi N \tau g^2}} e^{-g^2NT/2\tau} \frac{\pi^2}{48\tau^2}
\end{align}
We can similarly define a parameter $\epsilon$ with $I_\text{MLE} = (1-\epsilon) \, I_C$, $\epsilon \ll 1$, $I_C = 8g^2NT^3/3\tau + 4NT^2$, and $I_\text{MLE} = 1/\text{var(MLE)}$. Then, $\epsilon$ is 
\begin{subequations}
\begin{align}
    \epsilon = 1-\left[ 1 - q  + q \frac{\pi^2 N}{12} \left( \frac{T}{\tau}\right)^2 \left( 1 + \frac{2g^2T}{3\tau}\right)\right]^{-1} &\approx q \frac{\pi^2 N}{12} \left( \frac{T}{\tau}\right)^2 \left( 1 + \frac{2g^2T}{3\tau}\right) \\
    &= \frac{\pi^{3/2}}{48} N \left( \frac{T}{\tau} \right)^3 \sqrt{\frac{2\tau}{g^2 N T}} \left( 1 + \frac{2g^2T}{3\tau}\right) e^{-g^2NT/2\tau}
\end{align}
\end{subequations}
where $1-q$ is the probability of having no outliers in Eq. (\ref{eq:prob_no_outliers_final}). Defining $x = g^2NT/2\tau$, we obtain the following transcendental equation:
\begin{align}
\label{eq:transcendental_2}
    e^x \approx \frac{\pi^{3/2}}{48\epsilon} \frac{1}{\sqrt{x}} \left(\frac{T}{\tau}\right)^3 \left( N + \frac{4x}{3} \right)
\end{align}
Taking the logarithm on both sides, we obtain on the first order
\begin{subequations}
\begin{align}
    x^* &\approx \ln{\left(\frac{\pi^{3/2}}{48\epsilon} N \left(\frac{T}{\tau}\right)^3\right)} + \frac{4 x^*}{3 N} \\
    x^* &\approx \left[1- \frac{4}{3N}\right]^{-1} \ln{\left(\frac{\pi^{3/2}}{48\epsilon} N \left(\frac{T}{\tau}\right)^3\right)} \label{eq:x*_defn_Strong_meas}
\end{align}
\end{subequations}
Here, we assume that $4x^*/3N < 1$, i.e. $g^2T/\tau < 3/2$, in order to write $\ln{(N + 4x/3)} \approx \ln{(N)} + 4x/3N$. In order to find the second order approximation to the solution, let us write $x = x^* + \Delta x$. We obtain:
\begin{align}
    x^* + \Delta x \approx \ln{\left(\frac{\pi^{3/2}}{48\epsilon} N \left(\frac{T}{\tau}\right)^3\right)} -\frac{1}{2} \ln{(x^* + \Delta x)} + \frac{4}{3N}(x^*+\Delta x)
\end{align}
Plugging in the definition of $x^*$ in Eq. (\ref{eq:x*_defn_Strong_meas}), we finally write that $\Delta x \approx -\left[1- 4/3N\right]^{-1} \ln{(x^*)}/2$. Combining the first and second order terms, we find that the second order approximation to the solution to Eq. (\ref{eq:transcendental_2}) is
\begin{subequations}
\label{eq:threshold_cond_2}
\begin{align}
    \frac{g^2 N T}{2\tau} &\approx \left[1- \frac{4}{3N}\right]^{-1} \left[ \ln{\left(\frac{\pi^{3/2}}{48\epsilon} N \left(\frac{T}{\tau}\right)^3\right)} - \frac{1}{2} \ln{\left(\ln{\left(\frac{\pi^{3/2}}{48\epsilon} N \left(\frac{T}{\tau}\right)^3\right)} \right)} +\frac{1}{2}\ln{\left(1-\frac{4}{3N} \right)} \right] \\
    &\approx \ln{\left(\frac{\pi^{3/2}}{48\epsilon} N \left(\frac{T}{\tau}\right)^3\right)} - \frac{1}{2} \ln{\left(\ln{\left(\frac{\pi^{3/2}}{48\epsilon} N \left(\frac{T}{\tau}\right)^3\right)} \right)} 
\end{align}
\end{subequations}
where in the second line, $1-4/3N \approx 1$ as $N \gg 1$. Different from the previous case, in order to maximize the information gained from the measurement with respect to the measurement strength, $g$, we need to operate with the smallest possible $g$ (see Eq. (\ref{eq:fi_with_decay_strong})). Assuming that we fix $g T/\tau$ in Eq. (\ref{eq:threshold_cond_2}), we see that the necessary number of qubits to surpass the threshold and get close to the CFI scales again as $N \sim O(3 \ln{(T/\tau)})$.

In Fig. \ref{fig:mse_wrt_T}, we plot the sensitivity obtained by the weak measurement protocols, quantified by the inverse of the BMSE $(\Delta\widetilde{\omega})^2$, scaled by the QFI $I_Q$. We observe that the sensitivity approaches the CFI $I_C^\text{w}$, $I_C^\text{ws}$ in the weak back action regime $g^2T/\tau < 1$ for $N >20$. Then, we can find where the sensitivity reaches the CFI in this regime when $g^2NT/\tau \gg 1$, for $N > 20$ using Eqs. (\ref{eq:threshold_cond_1}) and (\ref{eq:threshold_cond_2}).

\section{Discussion on other types of prior distributions}

For a simpler analysis, we have assumed a uniform prior distribution in $[0, \delta\omega]$ in the main text. However, our protocol can be extended to other relevant prior distributions, such as a Gaussian distribution: $\mathcal{P}_{\delta\omega}\left(\omega\right)=\frac{1}{\sqrt{2\pi\left(\delta\omega\right)^{2}}}\exp\left(-\frac{\left(\omega-\omega_{0}\right)^{2}}{2\left(\delta\omega\right)^{2}}\right),$ 
where $\omega_0$
is the mean and the standard deviation is $\sigma = \delta\omega$. Let us comment on how the analysis would be modified to accommodate such a prior distribution.

First of all, the prior distribution determines the measurement period $\tau$: given a uniform distribution in $[0, \delta\omega]$, the phase accumulated during time $\tau$ satisfies $\phi \in [0, 2\delta\omega\tau]$.
To avoid phase slips, this interval needs to be contained 
in $[0, \pi]$.
Hence $2\delta\omega\tau\leq\pi,$
and thus $\tau \leq \pi/(2\delta\omega).$
As a side note,
we remark that it may be possible to increase $\tau$ to $\pi/\delta\omega$
by using a dual quadrature readout \cite{shaw_multi-ensemble_2023} .
This requires measuring half of the sensors in $\sigma_{x}$ basis and the other half in $\sigma_{y}$ basis, i.e. applying a $R_z(\pi/2)$ to half of the atoms at the beginning of the interrogation. 
For prior distributions with infinite support the choice of $\tau$ is less trivial. $\tau$ needs to be small enough to ensure small enough probability of phase slips such that BMSE would not be affected.
Considering a Gaussian prior distribution,
let us first recenter the distribution such that the average accumulated phase is $\pi/2,$ i.e. $\phi_0:=2\omega_{0}\tau=\pi/2\Rightarrow\omega_{0}=\pi/\left(4\tau\right).$
A phase slip occurs if $2|\omega-\omega_{0}|\geq\pi/\left(2\tau\right).$
This then becomes an e.g. $4 \sigma$ event if $4\delta\omega=\pi/\left(4\tau\right).$ 
Let us denote the contribution to the BMSE from beyond-$\pi/\tau$ phase slips as 
$\Delta_{\text{slips}}^{2}:={\int_{\omega\notin\left[0,\pi/2\tau\right]}}\mathcal{P}_{\delta\omega}\left(\omega\right)(\Delta\omega)^{2}.$
We can upper bound this contribution in the e.g. $4\sigma$ case by 
\begin{align}
\Delta_{\text{slips}}^{2}\leq2\underset{8\delta\omega}{\overset{\infty}{\int}}\omega^{2}\mathcal{P}_{\delta\omega}\left(\omega\right)=2\frac{1}{\sqrt{2\pi\left(\delta\omega\right)^{2}}}\underset{8\delta\omega}{\overset{\infty}{\int}}\omega^{2}\exp\left(-\frac{\left(\omega-4\delta\omega\right)^{2}}{2\left(\delta\omega\right)^{2}}\right)\approx0.004\left(\delta\omega\right)^{2}.\end{align}
Hence, as long as $I_{C}^{-1}\geq0.004\left(\delta\omega\right)^{2}$ this contribution can be neglected and the threshold analysis should be similar to the uniform prior case.
Otherwise, 
a smaller $\tau$  should be chosen to reduce $\Delta_{\text{slips}}^{2}$.
Therefore, in the Gaussian case, larger $I_{C}$ and in particular larger N imply that smaller $\tau$ is needed.
After setting $\tau$, the threshold condition for the MLE can again be found by using Eq. (\ref{eq:var_mle}). However, the MLE in Eq. (\ref{eq:mle_1}) needs to be redefined to include the information from the new prior distribution: the relevant likelihood function is $p\left(x_{+}\left(n\right)|\omega\right)\mathcal{P}_{\delta\omega}\left(\omega\right).$
A detailed analysis of the threshold behavior in this case is left for future work.

\section{The cascaded scheme} 

Let us introduce the cascaded scheme \cite{rosenband2013exponential} for dynamic range extension and analyze its precision limits.
While in the standard Ramsey scheme, all the $N$ atoms evolve for a duration of $T$ and accumulate a relative phase of $2 \omega T,$
in this method we partition the $N$ qubits into $M$ ensembles of $N'=N/M$ qubits that evolve for durations of $T,T/2,...,T/2^{M-1}.$
The state given this cascaded scheme can be thus written as $|\psi\rangle=2^{-N/2}{\Pi}_{j=0}^{M-1}(|0\rangle+|1\rangle e^{-i2\omega T/2^{j}})^{N'}.$
We refer to the $M-1$ ensembles that evolve for $T/2,...,T/2^{M-1}$ as blocks of slow atoms. These states have been analyzed in the literature extensively \cite{direkci2024heisenberg, rosenband2013exponential, kessler_heisenberg-limited_2014}.
The performance of this scheme is quite limited: it does not saturate the ultimate QFI bound of $4NT^2$, not even in the asymptotic limit, due to the equipartition of qubits in the different ensembles.
To show this, it suffices to compute the QFI obtainable with this protocol as a function of $N'$, $M$, and $T$,
and show that it does not reach this limit.
Since the initial state is a product state, the total QFI is the sum of the QFI's of all of the qubits. 
The state of a qubit given an interrogation time $\tau$ is $1/\sqrt{2}\left(|0\rangle+|1\rangle e^{-i2\omega \tau}\right),$ and thus 
its QFI is given by $4 \tau^2.$
The QFI of the ensemble that contains $N'$ qubits is then $4N'\tau^2.$
Summing over the QFI of all ensembles with interrogation times $T, T/2, \dots, T/2^{M-1}$, we compute the total QFI to be
\begin{align}
    I_B = 4 N' T^2 \sum_{i = 0}^{M-1} \left( \frac{1}{4} \right)^i = \frac{16 NT^2}{3M} \left(1 - \frac{1}{4^M}\right) < 4NT^2 \text{ for } M > 1.
\end{align}
Therefore, the total QFI is always smaller than the ultimate limit of sensitivity achievable with classical interrogations, $4NT^2$, if we have more than one ensemble.
In the limit of large prior width, $\delta \omega T\gg 1,$ ensembles of slow atoms are necessary and the minimal required $M$ is $M \approx \log \left(\delta\omega T \right)$ \cite{rosenband2013exponential,kessler_heisenberg-limited_2014}, as we need to have $\delta\omega T/2^{M-1}<\pi$ to prevent phase slip errors.
The intuition is that the $k$-th ensemble provides an estimate to the $k$-th binary digit of $2\omega T/\pi.$ Therefore in the limit of large $\delta\omega T,$ the optimal $M$ is $M \approx \log \left(\delta\omega T \right)$
such that
\begin{align}
I_{B}^{\text{opt}}\approx\frac{16}{3\log\left(\delta\omega T\right)}NT^{2}.
\label{supp_eq:IB_opt}
\end{align}
We therefore lose a factor of $3 \log\left(\delta\omega T\right)/4$ irrespective of how large $N$ is.
This is in contrast to the proposed weak measurement protocol, which saturates $4NT^2$ with large enough $N.$
Furthermore, for small $N$ and large $\delta \omega T$ the performance of the cascaded protocol is 
even worse: if $N/\log(\delta \omega T)$ is not large enough, then the relevant QFI, $I_B$, is not saturable. Taking e.g. $N=64,$ as in Fig. \ref{fig:threshold_mle_victory}b in the main text, and $\delta \omega T=100,$ then $M=7,$ and $N'\approx 9.$ This $N'$ is not large enough to attain $I_B$.

\begin{figure}
    \centering
\includegraphics[width=0.5\linewidth]{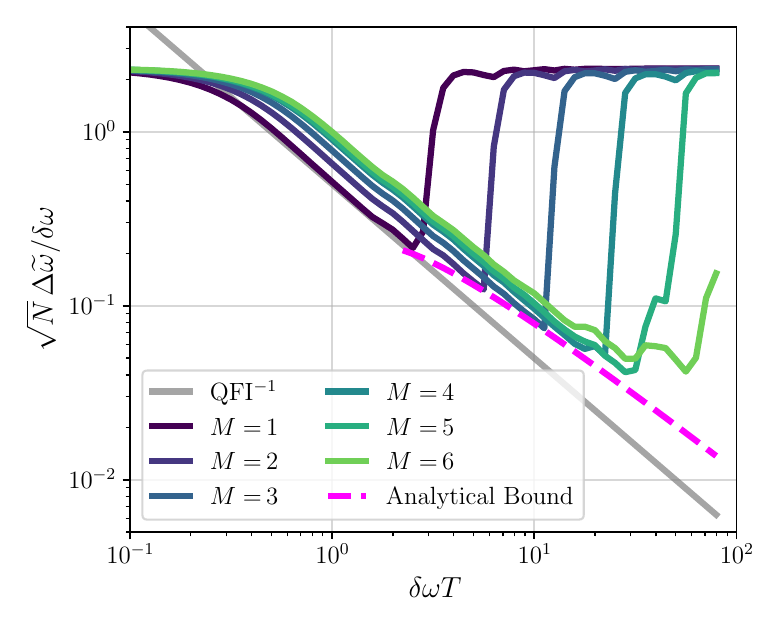}
    \caption{Square root of the BMSE of the cascaded protocol, $\Delta\widetilde{\omega}$, normalized by the prior width $\delta\omega$, as a function of the interrogation time $T$, for different number of ensembles $M$. We assume a uniform prior distribution in $[0, \delta\omega]$, and $N=64$ qubits. The overall performance of the cascaded protocol will be the minimum of $\Delta\widetilde{\omega}$ over all ensembles, as plotted in Fig. 3b in the main text. We compare the performance to i) the ultimate precision bound, given by $4NT^2$ and plotted with the gray curve, ii) the analytical approximation of $I_B^{\text{opt}}$ given in Eq. (\ref{supp_eq:IB_opt}) plotted with the magenta curve. The cascaded protocol diverges from the $I_B^{\text{opt}}$ for $M > 4$, as the number of atoms in each ensemble becomes too small to prevent phase slip errors. The BMSE increases beyond this ensemble number.
    }
\label{fig:rosenband_ensembles}
\end{figure}

We plot the performance of the cascaded protocol as a function of the interrogation time $T$ for ensemble numbers $M= 1, 2, \dots, 6$, as well as the analytical approximation of $I_{B}^{\text{opt}}$ in the limit of large $\delta\omega T$ (Eq. (\ref{supp_eq:IB_opt})) in Fig. \ref{fig:rosenband_ensembles}. We observe that for $M >4$, the sensitivity starts diverging from $I_{B}^{\text{opt}}$, as the number of qubits in each ensemble is not large enough to prevent phase slips.

\section{Imperfect Measurements}

Let us consider the following noisy ancilla measurement, in which the ancilla is measured through the following positive operator-valued measure (POVM):
\begin{align}
\label{eq:imperfect_povm}
 M_{0}=\left(1-p_{e}\right)|0\rangle\langle0|+p_{e}|1\rangle\langle1|, \;\; M_{1}=\left(1-p_{e}\right)|1\rangle\langle1|+p_{e}|0\rangle\langle0|,     
\end{align}
instead of the perfect projective measurement POVM of $\{\Pi_{0}=|0\rangle\langle0|, \Pi_{1}=|1\rangle\langle1|\}.$
This noisy measurement can be due to a symmetric bit-flip noise inflicted on the ancilla followed by a perfect projective measurement, or due to an error in the detection process itself. The probability of a bit-flip is parametrized through $p_e$.
The sensor and ancilla density matrix following this bit-flip noise is given by:
\begin{align}
    \Lambda(\rho_s \otimes \rho_a) = \left( (1-p_e) K_+ \rho K_+^\dagger + p_e K_- \rho K_-^\dagger \right) \otimes \ketbra{1} + \left( (1-p_e) K_- \rho K_-^\dagger + p_e K_+ \rho K_+^\dagger \right) \otimes \ketbra{0}
\end{align}
where $K_{\pm}$ are the weak measurement Kraus operators given in Eq. (\ref{eq:kraus_ops}). Hence, the state of the sensor after this measurement is updated as follows:
\begin{align}
\rho\rightarrow\frac{1}{p\left(x|\omega\right)}\left[\left(1-p_{e}\right)K_{(-1)^x}\rho K_{(-1)^x}^{\dagger}+p_{e}K_{(-1)^{x+1}}\rho K_{(-1)^{x+1}}^\dagger\right].    
\end{align}
$p(x|\omega)$ correspond to the probabilities of the measurement results, and $x=0,1$. In order to further understand the effect of such a channel on the sensing qubit, let us denote $\rho = 1/2 \, (I + \vec{r}\cdot \vec{\sigma})$, where $\vec{r} = (r_x, r_y, r_z)$. Furthermore, let us define $\phi = \tan^{-1}{(r_y/r_x)}$. 
The measurement probabilities after the ancilla readout will be given by
\begin{align}
\label{eq:imperfect_meas_probs}
    p(x|\omega) &= \frac{1}{2} \left[1 + (-1)^x (1-2 p_e) \sin{(2g)} \left( r_x \cos{(2\omega \tau)} + r_y \sin{(2\omega \tau)}\right)\right] \nonumber \\
    &= \frac{1}{2} \left[1 + (-1)^x (1-2 p_e) \sin{(2g)} r \cos{(2\omega \tau - \phi)}\right],
\end{align}
where we defined $r = |\vec{r}|$. The update on the parameters $r, \phi$ after the measurement are calculated as
\begin{subequations}
\label{eq:imperfect_update}
\begin{align}
    r^2 &\rightarrow \frac{r^2(1-\sin{(2g)}^2 \sin{(\phi-2\omega\tau)}^2) + (-1)^x 2 (1-2\,p_e) \sin{(2g)} \, r\cos{(\phi-2\omega\tau)}+(1-2\,p_e)^2 \sin{(2g)}^2}{\left(1 + (-1)^x \sin{(2g)}(1-2\,p_e) r \cos{(\phi-2\omega\tau)}\right)^2}, \\
\phi&\rightarrow\text{arg}\left(r\cos{(\phi-2\omega\tau)}+ (-1)^x(1-2\,p_{e})\sin{(2g)}+ir\cos{(2g)}\sin{(\phi-2\omega\tau)}\right).
\end{align}
\end{subequations}
$r$ and $\phi$ are initialized at the beginning of interrogation to $r = 1$, $\phi = 0$, as the sensing qubit is in the state $\ket{+} = (\ket{0}+\ket{1})/\sqrt{2}$.
For a perfect measurement, $p_e = 0$, $r$ is unchanged throughout the interrogation, as the sensing qubit remains in a pure state. However, for $p_e \neq 0$, the sensing qubit becomes a mixed state due to imperfect measurements, and $r$ undergoes a stochastic process (in addition to $\phi$).

\begin{figure}
    \centering
\includegraphics[width=0.95\linewidth]{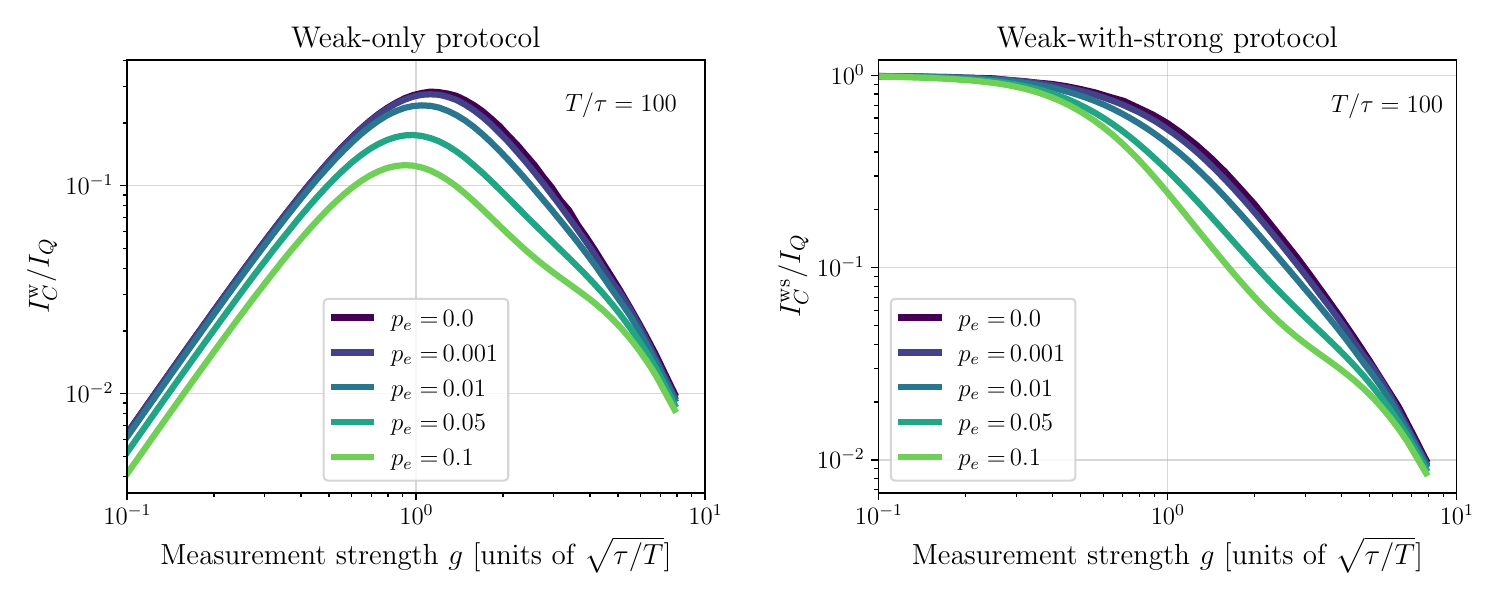}
    \caption{Effect of imperfect weak measurements on the CFI, with respect to the measurement strength $g$. We plot the CFI $I_C$ normalized by the QFI $I_Q$ as a function of $g$ and for different error probabilities $p_e$.  The left plot corresponds to the weak-only protocol, and the right one to the weak-with-strong protocol. We fix the interrogation time, the measurement period, and the frequency-to-be-estimated as $T = 10$ s, $\tau = 0.1$ s, and $\omega = 5$ rad, respectively. Note that in the weak-with-strong protocol (plotted on the right) we assume that the strong projective measurement is noiseless.}
    \label{fig:fi_wrt_g_imperfect}
\end{figure}

In what follows we derive the noisy CFI in the weak back action regime. An analytical derivation of the CFI in the strong back action regime is left for future work. In the weak back action limit, Eqs. (\ref{eq:imperfect_update}) can be approximated as
\begin{subequations}
\begin{align}
    r&\rightarrow r+ (-1)^x2g\left(1-2p_{e}\right)r\left(1-r^{2}\right)\cos\left(\phi - 2\omega\tau\right), \\
    \phi &\rightarrow \phi - 2\omega\tau - (-1)^{x} 2g (1-2p_e) \sin{(\phi - 2\omega\tau)}.
\end{align}
\end{subequations}
Then, we observe that up to first order in $g$, an initial $r=1$ (pure state) will remain equal to $1$ throughout the interrogation. Combined with the expression for the measurement probabilities in Eq. (\ref{eq:imperfect_meas_probs}), it can be seen that in this limit, the imperfect measurement effectively rescales the measurement strength $g$ as $g \rightarrow g \, (1-2p_e)$. Therefore, the CFI of the weak-only protocol in this limit is $8 Ng^2 (1-2p_e)^2 T^3/3\tau$. Regarding the CFI with the weak-with-strong protocol in this regime, it can be seen that CFI remains $4NT^2 - O(g^2)$ and the sensitivity gap compared to the noiseless case is negligible.
In the strong back action regime however the effect of $p_e$ is different: the gap from the noiseless CFI becomes larger and we observe that the scaling with time is no longer $T$ but modified to $T^{\alpha}$ with $\alpha<1.$ In the strong back action regime, the CFI of both weak measurement protocols converge.

\begin{figure}
    \centering
\includegraphics[width=0.5\linewidth]{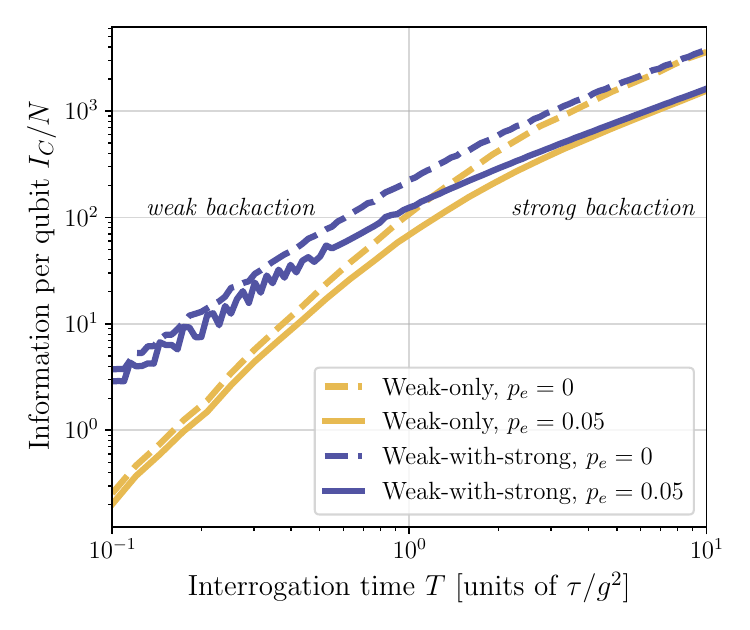}
    \caption{Effect of imperfect weak measurements on the CFI, with respect to the interrogation time $T$. We plot the CFI $I_C$ normalized by the number of qubits $N$ for the weak-only protocol and the weak-with-strong protocol with yellow and purple lines, respectively. We fix the measurement strength and period as $g=0.1$ and $\tau = 0.1$ s. Furthermore, we fix the frequency to be estimated as $\omega = 5$ rad/s. We observe that in the weak back action limit where $g^2T/\tau \ll 1$, the imperfect measurement effectively scales the measurement strength as $g \rightarrow g(1-2p_e)$, but the scaling with respect to $T$ is approximately unaffected such that the yellow dashed and plain curves are parallel to each other. In the strong back action limit, the CFI for the protocols with imperfect measurements (plain lines) converge to the classical $T$ scaling faster compared to the protocols with perfect measurements (dashed lines). We observed numerically that for larger $p_e$, the $T$ scaling gets modified to $T^\alpha$, $\alpha <1$.}
\label{fig:imperfect_meas_wrt_T}
\end{figure}

We plot the effect of imperfect measurements on the CFI in Figs. \ref{fig:fi_wrt_g_imperfect} and \ref{fig:imperfect_meas_wrt_T}.
In Fig. \ref{fig:fi_wrt_g_imperfect}, we plot the CFI as a function of $g$ for different measurement error values, $p_e$, for both protocols. As we found analytically, in the weak back action regime, the CFI with imperfect measurements is $I_C^{\text{w}}\approx 8 Ng^2 (1-2p_e)^2 T^3/3\tau$ for the weak-only protocol. Hence, the gap from the noiseless CFI in this limit is $\left(1-2p_{e}\right)^{2},$
and it is observed on Fig. \ref{fig:fi_wrt_g_imperfect}a. Furthermore, we observe that the CFI of the weak-with strong protocol is $I_C^{\text{ws}}\approx 4NT^2$ in the limit of small measurement strength $g \ll 1$.
In Fig. \ref{fig:imperfect_meas_wrt_T}, we plot the CFI as a function of the interrogation time $T$ and different values of $p_e.$ We notice that the CFI of both protocols converge to the same value in the strong back action regime, where the scaling with $T$ is modified due to imperfect measurements.

We can compute the threshold for which the information obtained from the MLE gets close to the CFI. For this purpose, we again work in the weak backaction regime $g^2T/\tau \ll 1$, and assume a large number of qubits and measurements ($N, T/\tau \gg 1$). From the discussion above, we can compute the thresholds given in Eq. (\ref{eq:threshold_cond_1}) and Eq. (\ref{eq:threshold_cond_2}) in this limit by rescaling $g \rightarrow g \,(1-2p_e)$.

To correct a finite number of bit-flip errors, we can prepare the ancilla qubits in an $n$-qubit GHZ state, i.e. in the state $\ket{\psi_{\text{GHZ}}}(\ket{1^n}+ i \ket{0^n})/\sqrt{2}$. Furthermore, one can engineer the following unitary interaction between the sensing qubit and the ancilla state: $U = \exp( - i g \, \sigma_x \otimes \sum_i \sigma_x^i)$, where the summation of the Pauli $x$ operators is over the ancilla qubits. After the free evolution and this unitary, the joint state of the sensing qubit and the ancillae $\rho_s \otimes \rho_a$ will be the following:
\begin{align}
    \Lambda(\rho_s \otimes \rho_a) = K_+ \rho_s K_+^\dagger \otimes \ket{1}^{\otimes n} \bra{1}^{\otimes n} + i K_- \rho_s K_+^\dagger \otimes \ket{0}^{\otimes n} \bra{1}^{\otimes n} - i K_+ \rho_s K_-^\dagger \otimes \ket{1}^{\otimes n} \bra{0}^{\otimes n} + K_- \rho_s K_-^\dagger \otimes \ket{0}^{\otimes n} \bra{0}^{\otimes n}, 
\end{align}
where $K_{\pm}$ are the weak measurement Kraus operators given in Eq. (\ref{eq:kraus_ops}), and $\rho_s$ is the state of the sensing qubit. A noisy measurement on the ancilla qubits will be described by the POVM in Eq. (\ref{eq:imperfect_povm}). Let us act on this state with the POVM element $M_0^m M_1^{n-m}$. The probability of obtaining this element will be given by $\text{Tr}[\Lambda(\rho_s \otimes \rho_a) I \otimes M_0^m M_1^{n-m}] = p_e^{m}(1- p_e)^{n-m} \text{Tr}[K_+ \rho_s K_{+}^\dagger] + (1-p_e)^{m} p_e^{n-m} \text{Tr}[K_- \rho_s K_{-}^\dagger]$. Furthermore, after tracing out the ancilla qubits, the sensing qubit will be in the state of
\begin{align}
    \rho_s \rightarrow \frac{p_e^{m}(1- p_e)^{n-m} K_+ \rho_s K_{+}^\dagger + (1-p_e)^{m} p_e^{n-m} K_- \rho_s K_{-}^\dagger}{p_e^{m}(1- p_e)^{n-m} \text{Tr}[K_+ \rho_s K_{+}^\dagger] + (1-p_e)^{m} p_e^{n-m} \text{Tr}[K_- \rho_s K_{-}^\dagger]}
\end{align}
For a very weak measurement, $\text{Tr}[K_- \rho_s K_{-}^\dagger] \approx \text{Tr}[K_+ \rho_s K_{+}^\dagger]$. Let us also assume without loss of generality that $m \ll n$, $n \gg 1$. Then, the probability of the state of the sensing qubit being $K_- \rho_s K_{-}^\dagger$ after obtaining the POVM element $M_0^m M_1^{n-m}$ is on the order of
\begin{align}
    \frac{p_e^{n-2m}}{p_e^{n-2m} + (1-p_e)^{n-2m} } \approx \left(\frac{p_e}{1-p_e}\right)^{n-2m}
\end{align}
Therefore, the conditional error probability is suppressed with $(p_e/(1-p_e))^n$.

\section{Possible extensions to entangled states and Heisenberg limit with $N$}

\begin{figure}[h!]
    \centering
\includegraphics[width=0.5\linewidth]{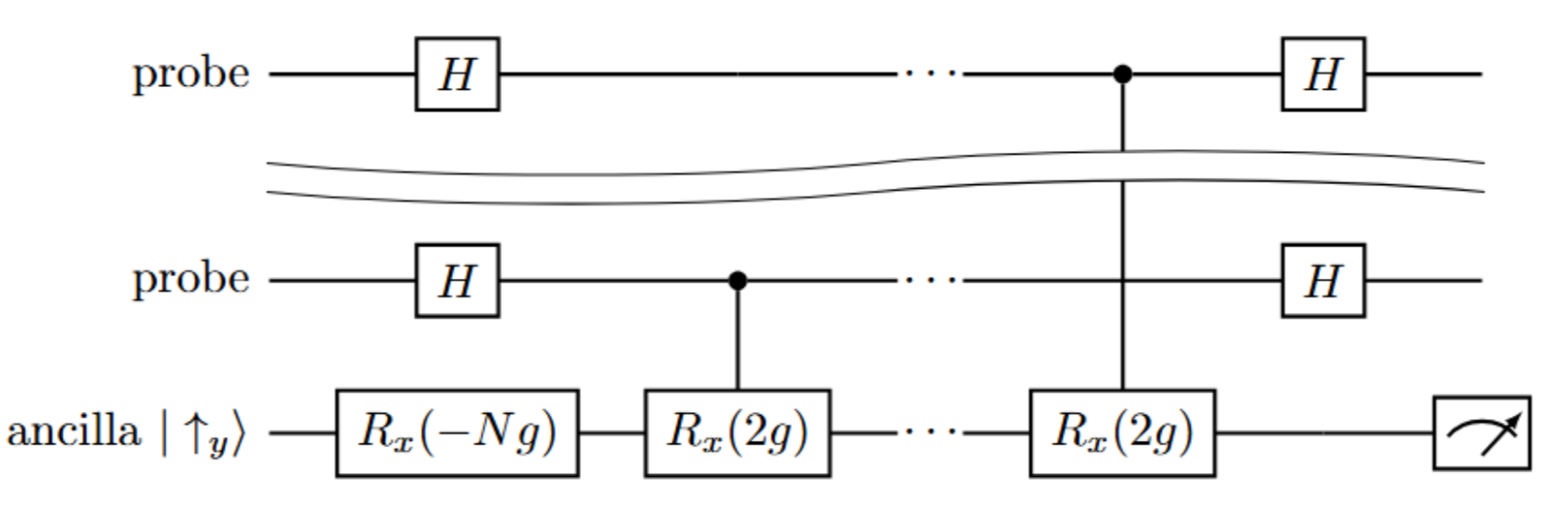}
    \caption{An illustration of a quantum circuit for the implementation of a binary version of the collective weak measurement of $J_{x}.$}
\label{fig:collective_weak_meas}
\end{figure}

We outline a possible extension of our weak measurement scheme to the sub-standard quantum limit (SQL) regime. This can be done by replacing the individual weak measurements with a collective weak measurement of the sensors. Denoting the total spin operators of the $N$ sensors as $J_{k}=\frac{1}{2}\sum_{i=1}^N \sigma_{k,i}$, where $k\in\left\{ x,y,z\right\}$, we consider weak measurements of $J_{x}$. This weak measurement, or a modified version of it, can be implemented by entangling all $N$ qubits to a single ancilla via sequential conditional rotations followed by measurement of the ancilla, as illustrated in Fig. \ref{fig:collective_weak_meas}.
These collective weak measurements can be then used to both generate a squeezed state and monitor its rotation. A similar protocol was previously considered in Refs. \cite{albarelli2017ultimate,tratzmiller2020limited, rossi_noisy_2020}, where it was shown that these weak measurements can generate spin squeezing that allows for a $N^2$ scaling of the QFI.
It is thus expected that given a large enough SNR, such a scheme can achieve HS in the BMSE, in the large bandwidth regime. We further expect that the necessary SNR can be obtained with a large enough $N$ and number of ancillas. A detailed analysis of this scheme is left for future work.

\section{Weak Measurements with Light}

Here, we consider a related but separate system, where the atoms are stored in a cavity. In this system, weak measurements can be performed by coupling a cavity mode dispersively to the atomic spin \cite{bowden_improving_2020, borregaard_near-heisenberg-limited_2013}. Such a coupling results in the  interaction Hamiltonian $\mathcal{H}_\text{int} = \chi \hat J_x \hat X$, where $\hat X$ is the position operator of the light,  field, $\hat J_x$ is the $x$ component of the total angular momentum operator $\hat{\mathbf{J}}$ of the atoms, and $\chi$ is some interaction strength. $\chi$ is related to the measurement strength $g$ in the proposed protocol. After the interaction, homodyne measurement is performed on the momentum operator $\hat P$ of light field to  probe the atomic spin. Let us assume that the atoms are initialized in a coherent state $\ket{+} = \left[(\ket{0} + \ket{1})/\sqrt{2}\right]^{\otimes N}$, where $N$ is the number of atoms. Therefore, we initially have: $\langle \hat J_y \rangle = \langle \hat J_z \rangle = 0$, $\langle \hat J_x \rangle = N/2$, $\langle \hat J_y^2 \rangle = \langle \hat J_z^2 \rangle = N/4$, and $\langle \hat J_x^2 \rangle = N^2/4$, such that $\Delta \hat J_y =\Delta \hat J_z = \sqrt{N/4}$, and $\Delta \hat J_x = 0$.

The free evolution can be described by a rotation of the total angular momentum vector on the collective Bloch sphere, quantified by a rotation matrix $\mathbf{U}(\omega\tau)$, where $\omega$ is the frequency to be estimated and $\tau$ is the free evolution time. A weak measurement is performed immediately after the free evolution, where the outgoing momentum operator of the light field is given by
\begin{align}
    \hat P_\text{out} = \hat P_\text{in} - \chi t' \, \hat J_x.
\end{align}
where $t'$ is the weak measurement time. We assume that the probing light has vacuum statistics at any time, such that $\langle\hat X_\text{in}\rangle = \langle\hat P_\text{in}\rangle = \langle\hat X_\text{in}P_\text{in}\rangle = 0$, and $\langle\hat X_\text{in}^2\rangle = \langle\hat P_\text{in}^2\rangle = 1/2$. The weak measurement will cause a back action on the collective angular momentum of the atoms, which also can be described by a rotation on the Bloch sphere, quantified by a rotation matrix $\mathbf{M}(\hat \Pi)$, with $\hat \Pi = \chi t' \hat X$. We describe the collective dynamics of the atoms using the Heisenberg picture of the angular momentum operators $\hat{\mathbf{J}}(t)$, $t>0$. The effect of one cycle of free evolution and weak measurement can be written as 
\begin{align}
\label{eq:light_first_moment}
     \hat{\mathbf{J}}(t+\tau)
     = \mathbf{M}(\hat \Pi) \mathbf{U}(\omega \tau)
     \hat{\mathbf{J}}(\tau), \;\; \mathbf{U}(\omega \tau) = \begin{bmatrix}
         \cos{(2\omega \tau)} & -\sin{(2\omega \tau)} & 0 \\ \sin{(2\omega \tau)} &  \;\;\; \cos{(2\omega \tau)}  & 0 \\ 0 & 0 & 1
    \end{bmatrix}, \;\; \mathbf{M}(\hat \Pi) = \begin{bmatrix}
        1 & 0 & 0 \\  0 & \cos{(\hat \Pi)} & -\sin{(\hat \Pi)} \\ 0 & \sin{(\hat \Pi)} & \;\;\;\, \cos{(\hat \Pi)}
    \end{bmatrix}  
\end{align}
In order to find the sensitivity with respect to $\omega$ at any given time $t$, we can calculate
\begin{align}
    (\Delta \omega)^2 \approx \frac{(\Delta \hat J_x(t))^2}{(\partial\langle \hat J_x(t) \rangle/ \partial\omega)^2}
\end{align}
Therefore, we want to find $J_x(T)$ at the end of the interrogation, where $T$ is the interrogation time. In order to compare a protocol that uses the light field to perform weak measurements to our protocol, we assume that weak measurements with light are performed periodically (with a period of $\tau$), and a projective measurement on the atomic ensemble is performed at the end of the interrogation. Furthermore, we aim to take the limit of infinitesimally weak measurements $\chi t' \rightarrow 0$, in order to observe if this protocol also reaches the QFI limit of $4NT^2$.

First, to calculate the first order moment of $\hat J_x$, we can write from Eq. (\ref{eq:light_first_moment})
\begin{align}
\label{eq:moment_analysis}
    \langle
     \hat{\mathbf{J}}(n\tau)
    \rangle = \prod_{i=1}^n \langle\mathbf{M}_i(\hat \Pi) \rangle \mathbf{U}(\omega \tau) \langle
     \hat{\mathbf{J}}(0)
    \rangle
\end{align}
where $\mathbf{M}_i(\hat \Pi)$ is the back action matrix from the $i^\text{th}$ weak measurement. In order to find $\langle\mathbf{M}_i(\hat \Pi) \rangle$, we observe that for linear operators on Gaussian states, we have: $\langle e^{\hat A} \rangle = e^{\langle \hat A \rangle} e^{(\Delta \hat A )^2/2}$. Then, $\langle \cos{(\hat \Pi)}\rangle = \cos{((\chi t'/2)^2)}$, $\langle \sin{(\hat \Pi)}\rangle = \sin{((\chi t'/2)^2)}$. To make the analysis simpler, we can add another rotation matrix after each weak measurement to cancel this average rotation. Then, we will have at the end of the last measurement: $\langle\hat J_x(T) \rangle = \cos{(2\omega T)} \langle \hat J_x(0) \rangle - \sin{(2\omega T)} \langle \hat J_y(0) \rangle = \cos{(2\omega T)} N/2$. To find the second moments, we can do a similar analysis where we evolve the Heisenberg operators of $\{\hat J_{x,y,z}, \hat J_{x,y,z}\}$. We find in the limit that $T/\tau \gg 1$, $N \gg 1$
\begin{align}
    (\Delta \hat J_x(T))^2 = \frac{N}{4} \sin{(2\omega T)}^2 - \frac{(\chi t')^2}{64} N \frac{T}{\tau} \cot{(\omega\tau)} \sin{(4\omega T)}
\end{align}
Therefore, from Eq. (\ref{eq:moment_analysis}), we find that
\begin{align}
    \frac{(\Delta \hat J_x(T))^2}{(d\langle \hat J_x(T) \rangle/d\omega)^2} \approx \frac{1}{4NT^2}
\end{align}
in the limit that $\chi t' \rightarrow 0$. Therefore, we observe that $(\Delta\omega)^2 \approx I_Q$ with this protocol. Then, the protocol that employs light to perform periodic weak measurements on the total angular momentum of an atomic ensemble, with a projective measurement at the end of the interrogation, reaches the ultimate sensitivity limit for coherent states in the limit of infinitesimally weak measurements. Note that the QFI, $I_Q$, is saturated asymptotically, i.e. in the limit of $N \rightarrow \infty$.

For non-zero measurement strengths $\chi t'$, the analysis gets significantly more complex, since weak measurements can squeeze the total angular momentum of the atomic ensemble, giving rise to non-classical scaling with respect to the number of atoms $N$ \cite{rossi_noisy_2020}. This behavior can be recreated in our protocol by coupling a single ancilla to multiple qubits (compare with the current protocol where a single qubit interacts with a single ancilla). We leave the details of such a protocol to future work.

\bibliography{refs_2}

\end{document}